\documentclass[journal,twoside,web]{ieeecolor}

\usepackage{generic}
\usepackage{cite}
\usepackage{amsmath,amssymb,amsfonts}
\usepackage{graphicx}
\usepackage{hyperref}
\hypersetup{hidelinks=true}
\usepackage{textcomp}
\usepackage{booktabs}
\usepackage{multirow}
\usepackage{array}
\usepackage{tabularx}
\usepackage{subcaption}
\usepackage{balance}
\usepackage{makecell}
\usepackage{xcolor}

\begin{document}

%\title{Efficient Cross-Dataset EEG Classification via\\Self-Supervised Learning and Knowledge Distillation}
\title{BRIDGE-EEG: Bridging Self-Supervised Pretraining and Efficient Deployment for Cross-Dataset EEG Classification}

%Authors: Meghna Roy Chowdhury, Chengwei Zhou, Haotian Yu, XXX, Gourav Datta, Shreyas Sen
%Chengwei Zhou, Haotian Yu, and Gourav Datta are with Case Western Reserve University

\author{Meghna Roy Chowdhury,~\IEEEmembership{Student Member,~IEEE,}
Chengwei Zhou,~\IEEEmembership{Student Member,~IEEE,}
Haotian~Yu,
Gourav Datta,~\IEEEmembership{Senior Member,~IEEE}
and~Shreyas Sen,~\IEEEmembership{Senior Member,~IEEE}
\thanks{ M. Roy Chowdhury and S. Sen are with Purdue University, Indiana, USA (e-mail: mroycho@purdue.edu, shreyas@purdue.edu) }
\thanks{Chengwei,Haotian and Gourav are with Case Western University, Ohio, USA (e-mail: cxz760@case.edu, haotian.yu@case.edu, gxd234@case.edu)}
}
\maketitle
~\vspace{-1.8cm}

\begin{abstract}

The growing use of electroencephalography (EEG) has created a need for automated analysis that is accurate and robust across diverse patients, tasks, and recording setups. Recent EEG foundation models learn transferable features from large-scale pretraining, but their size and computational cost make them difficult to deploy on edge and wearable hardware. Conversely, compact EEG models are easier to deploy but are usually trained from scratch for a single task, limiting generalization across datasets. We address this trade-off with BRIDGE-EEG, a pipeline for efficient multi-task EEG classification that preserves the benefits of pretraining while enabling deployment on constrained hardware. First, we introduce a unified preprocessing that maps heterogeneous EEG recordings with varying channel counts, montages, and sampling rates to a device-agnostic 62-channel time–frequency representation. Next, we pretrain an SE-ResNet18 teacher (11.84 M parameters) using SimCLR on unlabeled EEG from five heterogeneous datasets, then compress it into SE-ResNet8 (1.56 M) and SE-ResNet4 (0.48 M) students through task-agnostic and task-specific distillation strategies, respectively. We validate across six benchmarks spanning abnormality detection, motor imagery, and emotion recognition. On abnormality detection and emotion recognition, the students achieve accuracy comparable to or above several recent EEG foundation models with 10--1,000$\times$ more parameters. On motor imagery, a representation gap remains, pointing to pretraining diversity as a key factor. Finally, we profile inference on a server GPU, a desktop CPU, and an edge device (NVIDIA Jetson Orin Nano). where compression yields up to 3.0× lower energy per inference on edge hardware (15.64 mJ vs. 46.67 mJ). The resulting models' compact footprint further motivates deployment on MCU-class wearables.

% The compact size of the resulting models further suggests a viable path toward deployment on MCU-class wearable devices.

% outlining an empirical first step toward future wearable deployment.

% outlining an empirical first step toward future wearable deployment by demonstrating that edge-device-scale models can retain useful accuracy.

\end{abstract}

\vspace{-1mm}

%% ====================================================================
%%  I. INTRODUCTION
%% ====================================================================
\vspace{-4mm}
\section{Introduction}
\label{sec:introduction}

\begin{figure}[t]
    \centering
\vspace{-1cm}
\includegraphics[width=\columnwidth]{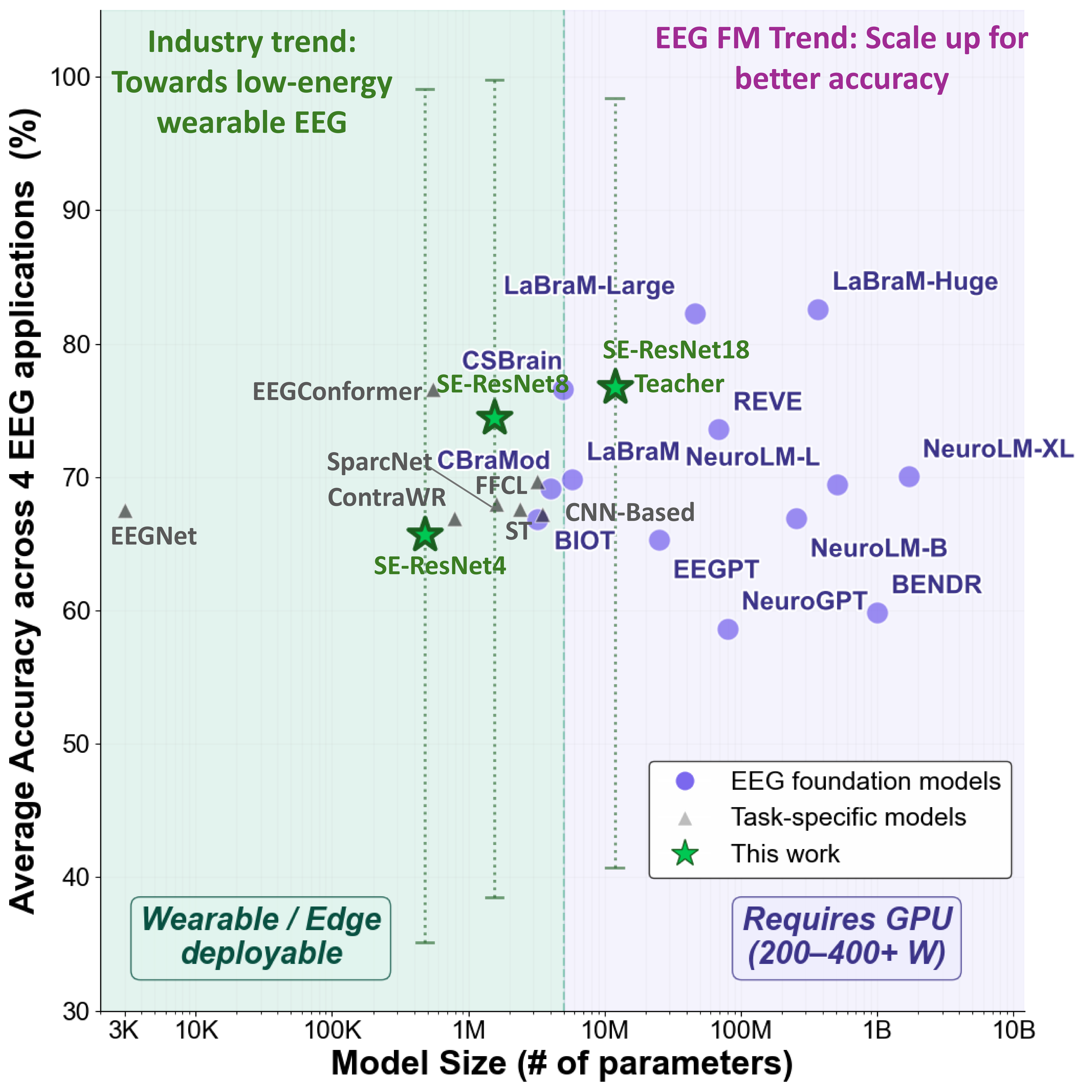}
% \caption{Positioning of this work. EEG foundation models offer broad transfer but limited efficiency characterization, whereas compact embedded models are efficient but typically task-specific. Our framework targets the middle ground: compact, transferable models profiled across three hardware power tiers.}
\caption{EEG model landscape: accuracy vs. parameter count. Our framework (stars) targets compact models profiled across three hardware tiers.}
\vspace{-7mm}
    \label{fig:landscape}
\end{figure}

Electroencephalography (EEG) is one of the most widely used modalities for non-invasive brain monitoring. It has a broad range of applications, such as seizure detection, neurological screening, and brain-computer interfaces~\cite{he2023diversity}. As these applications move from controlled laboratory studies toward continuous, real-world monitoring, there is a growing need for automated EEG analysis that is not only accurate but also robust across subjects, recording setups, and tasks, and ultimately feasible on resource-constrained devices. EEG is also difficult to model consistently at scale as recordings are often noisy, channel quality varies across sessions, electrode layouts differ across devices, and datasets are typically collected under isolated protocols. These challenges become more severe in real-world settings, where motion artifacts, missing electrodes, low-density montages, and hardware variability are common. As a result, many EEG learning pipelines remain tied to a single dataset, montage, or task, limiting their ability to generalize across realistic deployment conditions.

%  A broad pretrained model can therefore serve as a shared prior for downstream learning, improving robustness under heterogeneous recording conditions. Recently, there have been efforts to pursue this direction through large-scale EEG pretraining.

This motivates the use of foundation models (FM). Rather than training a separate network from scratch for each downstream dataset, FMs aim to learn reusable neural representations from large, diverse EEG corpora. This paradigm is especially helpful for EEG because the signal is information-rich but diverse across subjects and measurement hardware. In particular, EEG FMs~\cite{ouahidi2025reve,wei2025neurolm,kostas2021bendr,zhou2025csbrain,yang2023biot,wang2025cbramod,jiang2024labram} and MAMBA-based models ~\cite{panchavati2025mentality,wang2025eegmamba,tegon2025femba} aim to learn transferable representations from diverse EEG corpora and adapt them to multiple downstream applications. However, these models are primarily evaluated in terms of downstream accuracy, whereas their energy cost on resource-constrained hardware remains largely uncharacterized. In parallel, compact task-specific EEG models have also been investigated. Although these models are often easier to deploy, they are typically designed for a single dataset and task rather than derived from a shared pretrained backbone.

% In particular, recent EEG FMs, spanning both Transformer- and state-space-based architectures such as BENDR, LaBraM, NeuroLM, and FEMBA, aim to learn transferable representations from diverse EEG corpora and adapt them to multiple downstream applications.

This creates a gap between generalization and deployability. On one side, large pretrained models offer broader generalization across heterogeneous EEG data. On the other hand, compact embedded models are more practical for deployment but usually remain task-specific. Fig.~\ref{fig:landscape} illustrates this. This raises a key question: \textit{can the benefits of EEG foundation models be retained in a form that is sufficiently compact and efficient for deployment?} This work therefore investigates whether smaller models can retain accuracy comparable to that of EEG FMs, and characterizes the associated accuracy--energy trade-offs across three hardware tiers: a server-class GPU, a desktop CPU, and an edge device. This problem is nontrivial for several reasons. First, although large-scale EEG pretraining can improve transfer across datasets and tasks, it remains unclear how much of this transfer advantage can be retained after compression. Second, model size alone is not a sufficient measure of practical efficiency, as deployment ultimately depends on power consumption and energy. Third, the heterogeneity in EEG acquisition, including variations in channel count and sampling rate, makes it challenging for a single model to support diverse downstream tasks.

\begin{figure*}[t]
    \centering
    \includegraphics[width=1\linewidth]{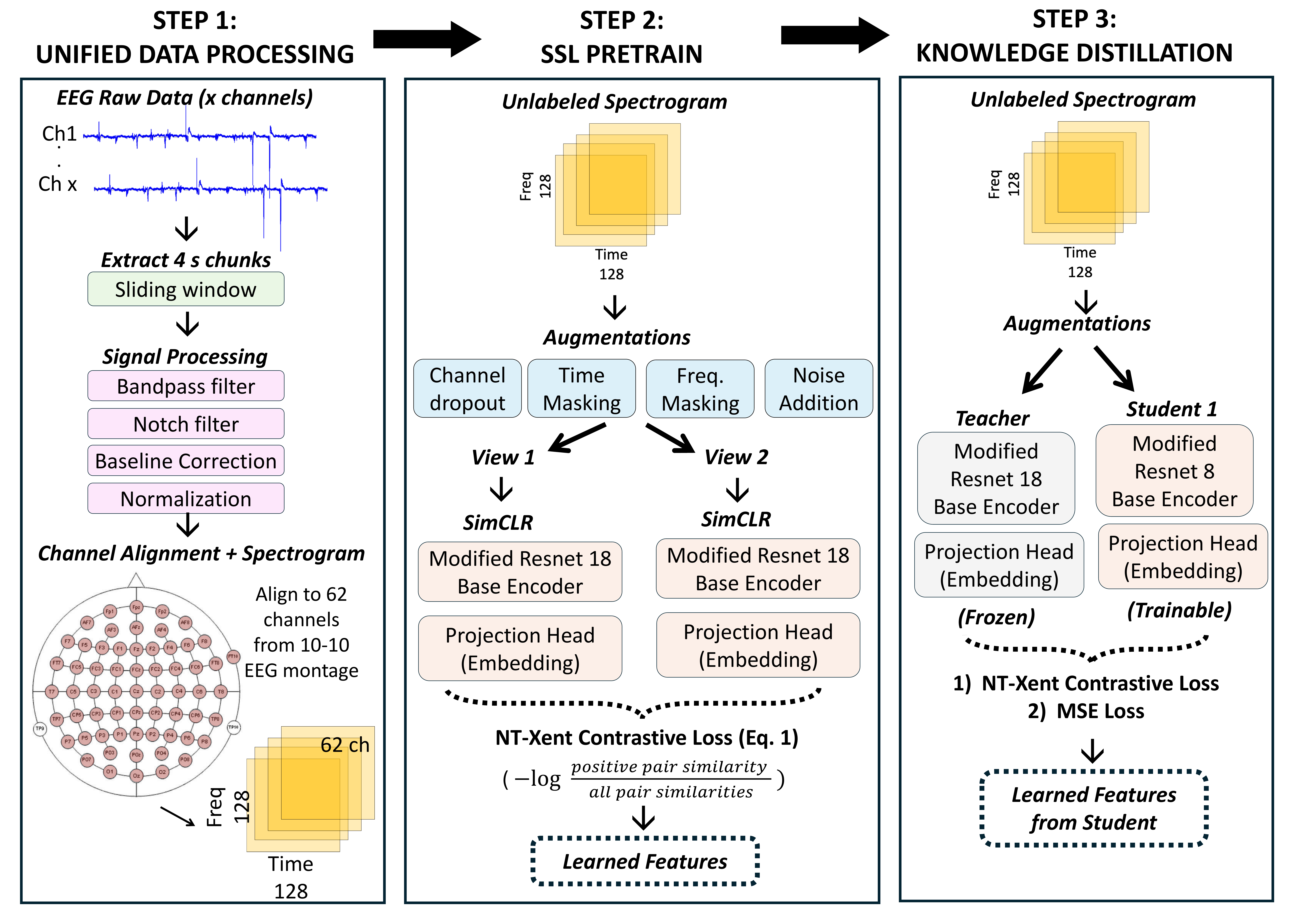}
    % \caption{Overview of BRIDGE-EEG. Step 1: heterogeneous EEG signals are aligned to a 62-channel STFT spectrogram representation. Step~2: an SE-ResNet18 teacher is pretrained via SimCLR on pooled unlabeled data. Step~3: Knowledge Distillation to compress the teacher into compact student.}
    \vspace{-8mm}
    \caption{BRIDGE-EEG pipeline. (1) Unified preprocessing to device-agnostic 62-channel STFT spectrograms. (2) SimCLR pretraining of SE-ResNet18 teacher. (3) Knowledge distillation into compact students.}
    \vspace{-5mm}
    \label{fig:pipeline}
\end{figure*}

To address these challenges, we propose BRIDGE-EEG, a three-stage pipeline for efficient multi-task EEG classification (Fig.~\ref{fig:pipeline}) that, to our knowledge, is the first EEG framework to combine self-supervised multi-dataset pretraining, post-pretraining knowledge distillation, and measured cross-tier inference profiling within a single pipeline. In the first stage, a unified preprocessing method maps heterogeneous EEG recordings into a device-agnostic 62-channel time–frequency representation, enabling a single shared backbone across fundamentally different acquisition setups. In the second stage, an SE-ResNet18 encoder is pretrained with SimCLR~\cite{chen2020simclr} using NT-Xent contrastive loss on pooled, unlabeled multi-dataset EEG, yielding a transferable teacher representation. In the third stage, the teacher is compressed via two KD strategies: task-agnostic distillation produces an SE-ResNet8 (1.56\,M) backbone that serves as a shared initialization across tasks, while task-specific distillation yields per-task SE-ResNet8 and SE-ResNet4 (0.48\,M) students that maximize individual task accuracy. The resulting model family achieves accuracy comparable to or above several recent EEG FMs on multiple benchmarks spanning three EEG application categories, with up to three orders of magnitude fewer parameters, while being measured systematically for latency and energy across server, desktop, and edge hardware tiers. 

This work extends our prior conference paper, SSL-SE-EEG~\cite{chowdhury2025sslse}, which introduced SimCLR pretraining with a squeeze-and-excitation encoder with 2D EEG image representations. The present study extends that foundation in scope, methodology, and evaluation. We move from a single SSL architecture study to a full SSL-to-KD pipeline, from limited downstream analysis to a multi-task evaluation, and from GPU-only latency and power profiling to measured energy-per-inference across three hardware tiers. The main contributions of this work are as follows:
\begin{enumerate}
    \item An \textbf{SSL-to-KD pipeline for efficient multi-task EEG classification}, in which an SE-ResNet18 teacher pretrained via SimCLR is compressed into SE-ResNet8 and SE-ResNet4 students using two distillation strategies.
    \item A \textbf{unified 62-channel spectrogram preprocessing
    pipeline} that maps heterogeneous EEG recordings to a common
    input representation, enabling a shared backbone across datasets acquired with different channel counts, montages, and sampling rates.
\item A \textbf{systematic evaluation of accuracy and efficiency across six downstream EEG applications and three hardware tiers (server-class GPU, desktop CPU, and NVIDIA Jetson Orin Nano edge device).} The compressed students' accuracy remains comparable to larger EEG FMs, while having edge-device efficiency. These results provide an empirical step toward future wearable EEG deployment.
\end{enumerate}

The remainder of this paper is organized as follows. Section~\ref{sec:related} reviews EEG FMs, self-supervised EEG pretraining, and knowledge distillation for efficient biosignal inference. Section~\ref{sec:framework} presents the proposed pipeline. Section~\ref{sec:setup} describes the datasets, training protocol, and hardware profiling setup. Section~\ref{sec:results} reports the empirical results across all downstream benchmarks and hardware tiers. Section~\ref{sec:discussion} discusses the implications for efficient and transferable EEG modeling, and Section~\ref{sec:conclusion} concludes the paper.

\section{Related Work}
\label{sec:related}

\begin{table}[!t]
    \centering
    \footnotesize
    \setlength{\tabcolsep}{2.5pt}
    \caption{SOTA EEG FMs.}
    \vspace{-1mm}
    \label{tab:fm-landscape}
    \begin{tabular}{l c p{1.0cm} l}
        \toprule
        \textbf{Model} & \textbf{Params} & \textbf{Hrs} & \textbf{SSL strategy} \\
        \midrule
        BENDR~\cite{kostas2021bendr}       & $>$1 B         & 21,000   & wav2vec SSL \\
        BrainBERT~\cite{wang2023brainbert} & --              & --          & masked pred. \\
        Brant~\cite{zhang2023brant}        & 505.69 M       & 2{,}528     & temporal-spatial \\
        BIOT~\cite{yang2023biot}           & 3.2 M          & 55,000       & contrastive loss \\
        LaBraM~\cite{jiang2024labram}      & 5.8--369 M     & 2{,}500     & VQ spectrum pred. \\
        NeuroGPT~\cite{cui2024neurogpt}    & 79.53 M        & 5{,}656     & autoregressive \\
        EEGPT~\cite{wang2024eegpt}          & 4.7--25 M     & --          & electrode-wise SSL \\
        NeuroLM~\cite{wei2025neurolm}      & 254--1{,}696 M & 25{,}000    & EEG tokens + GPT \\
        CBraMod~\cite{wang2025cbramod}     & 4.0 M          & 27{,}062    & masked recon. \\
        CSBrain~\cite{zhou2025csbrain}     & --              & $\sim$9{,}000 & masked autoencoding (MAE) \\
        REVE~\cite{ouahidi2025reve}        & 69--400 M & $\sim$60{,}000 & modified MAE \\
        % EEGFormer~\cite{wan2023eegformer}     & --              & $\sim$26{,}000 & -- \\
        \bottomrule
    \end{tabular}
    \vspace{-5mm}
\end{table}

% \subsection{EEG FMs}

\subsection{EEG Learning Models: From Task-Specific to FMs}

EEG modeling has evolved from task-specific supervised networks toward large-scale pretrained models that aim to learn transferable neural representations. Earlier task-specific approaches, such as EEGNet~\cite{lawhern2018eegnet}, ShallowConvNet, and DeepConvNet~\cite{schirrmeister2017deep}, as well as later attention-augmented and temporal-convolutional variants~\cite{hu2018senet,li2020tsseseizure,altuwaijri2022mbeegse,ingolfsson2020eegtcnet,song2023conformer,altaheri2023atcnet}, remain effective within individual applications, but they are typically trained from scratch on a single dataset and do not explicitly target cross-dataset generalization. Self-supervised and other pretraining objectives on large unlabeled EEG corpora have enabled a different paradigm in which reusable neural representations are learned before downstream fine-tuning. BENDR~\cite{kostas2021bendr} was one of the earliest prominent examples of this direction, showing that large-scale pretraining could support transfer across multiple EEG tasks. Since then, a growing family of EEG FMs has emerged, including BrainBERT~\cite{wang2023brainbert}, Brant~\cite{zhang2023brant}, BIOT~\cite{yang2023biot}, LaBraM~\cite{jiang2024labram}, NeuroGPT~\cite{cui2024neurogpt}, EEGPT~\cite{wang2024eegpt}, NeuroLM~\cite{wei2025neurolm}, and CBraMod~\cite{wang2025cbramod}. These models differ in architecture, tokenization strategy, and pretraining objective, but collectively reflect a shift toward large-scale EEG pretraining. More recently, state-space and Mamba-based EEG models have emerged as an efficient alternative to Transformer backbones~\cite{wang2025eegmamba,tegon2025femba,panchavati2025mentality}. In particular, FEMBA~\cite{tegon2025femba} extends large-scale self-supervised EEG pretraining to a bidirectional Mamba encoder trained on approximately 21{,}000 hours of TUEG, reporting competitive performance with reduced FLOPs and memory.
%  However, these works hardly report inference cost or deployment feasibility on constrained hardware.
In parallel, SSL methods tailored specifically to EEG have continued to mature outside the foundation-model label. Prior work explored temporal contrastive learning for clinical EEG~\cite{banville2021structure}, SimCLR-style augmentations for EEG~\cite{mohsenvand2020contrastive}, masked reconstruction objectives~\cite{chien2022maeeg,foumani2024eeg2rep}, and task-specific SSL for seizure detection and motor imagery~\cite{xiao2024slamseizure,li2024sslmi}. In our prior work~\cite{chowdhury2025sslse}, we combined SimCLR with squeeze-and-excitation encoders on EEG image representations and showed that SE attention improved performance with only modest power overhead. However, across most EEG FMs and SSL-based EEG pretraining studies, deployment efficiency remains less characterized, with existing work reporting theoretical FLOPs rather than measured on-device energy.

% the dominant emphasis remains representation quality and downstream accuracy rather than model compression or deployment efficiency.

% These models differ in architecture, tokenization strategy, and pretraining objective, but collectively they reflect a broader shift toward large-scale EEG pretraining and cross-task transfer.

\begin{figure*}[t]
    \centering
    \includegraphics[width=1\linewidth]{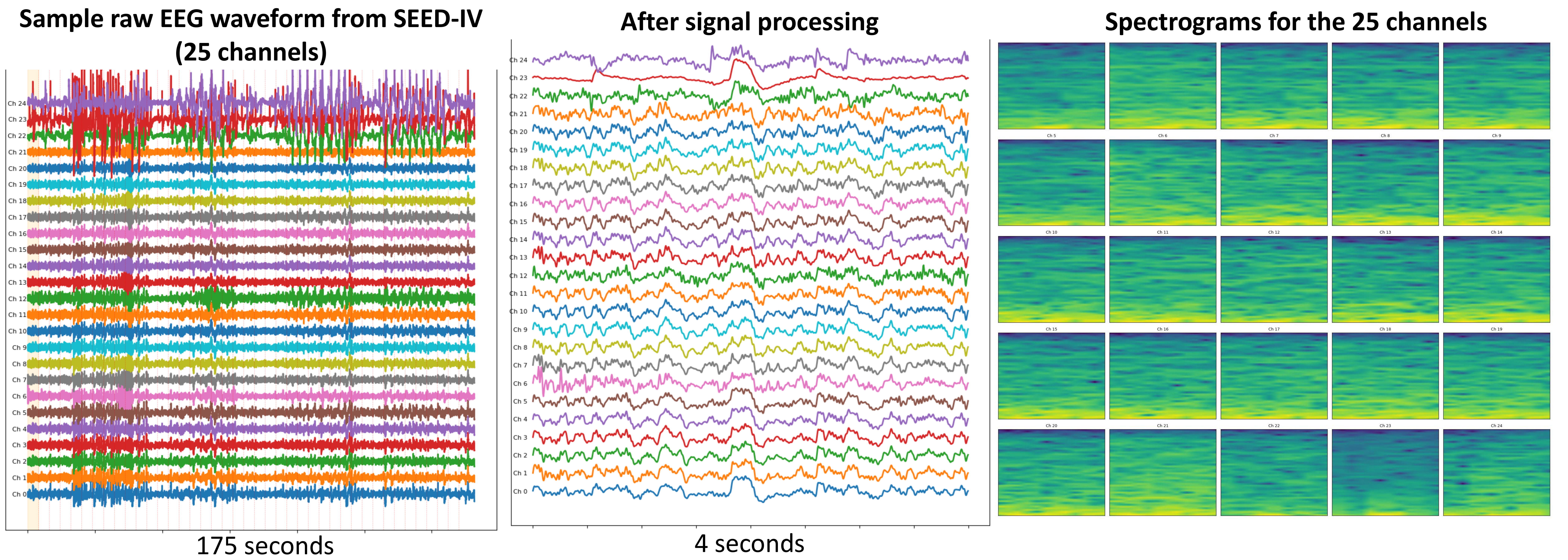}
    \caption{Preprocessing example from SEED-IV (25 channels shown here). Left: raw EEG (175 s). Middle: after filtering, baseline correction, and z-normalization (4 s window). Right: per-channel STFT spectrograms (128×128, 1–50 Hz).}
    \vspace{-5mm}
    \label{fig:preprocess}
\end{figure*}

\vspace{-2mm}
\subsection{Knowledge Distillation for Efficient Biosignal Inference}
Knowledge distillation~\cite{hinton2015distilling,romero2015fitnets} is a standard approach for compressing large models into smaller deployable ones, but its use in EEG has so far been narrow and mostly task-specific. Existing EEG distillation studies focus on channel reduction~\cite{huang2023skd}, sleep staging~\cite{liang2023sleepkd,jia2024distillsleep,zhang2022multichannel}, and emotion recognition~\cite{zhang2024llmkd,zhang2022visualeeg}. These works show that compact students can preserve useful task performance, but they do not begin from large self-supervised EEG backbones and generally do not report measured hardware efficiency. Related biosignal research has explored adjacent compression settings. Baghersalimi~et~al.~\cite{baghersalimi2024m2skd} used distillation to obtain a lower-power wearable seizure detector, Sepahvand~et~al.~\cite{sepahvand2022ecgkd} compressed 12-lead to single-lead ECG, and Abbaspourazad~et~al.~\cite{abbaspourazad2024wearablekd} demonstrated SSL-to-accelerometer distillation for wearable sensing. These studies highlight the promise of distillation for resource-constrained biosignal systems, but they do not address multi-task EEG compression from a shared pretrained backbone.  A separate line of work focuses directly on efficient EEG inference on embedded hardware. For instance, BrainFuseNet~\cite{ingolfsson2024brainfusenet} and EpiDeNet~\cite{ingolfsson2023epidenet} report low-energy inference on constrained processors, while MCU and FPGA/ASIC implementations~\cite{zanetti2020stm32,saric2020fpga,li2022asic65nm} push further toward ultra-low-power operation. Yet these systems are generally supervised, single-task, and designed directly for deployment, rather than distilled from a larger pretrained model.

\section{Proposed Framework}
\label{sec:framework}

Large-scale EEG pretraining improves cross-dataset transferability, while compact architectures and knowledge distillation enable resource-constrained deployment. However, these directions are typically examined in isolation, and their interactions across hardware tiers remain underexplored in multi-task EEG models. BRIDGE-EEG bridges this gap by combining self-supervised multi-dataset pretraining, knowledge distillation, and systematic inference profiling across three hardware tiers: a server-class GPU, a desktop CPU, and an NVIDIA Jetson Orin Nano edge device, achieving \textit{accuracy comparable to or above several recent EEG foundation models on abnormality detection and emotion recognition, while maintaining edge-class efficiency.} This outlines a practical pathway from high-performance foundation models to low-power wearable EEG systems. To this end, BRIDGE-EEG operates in three stages, as illustrated in Fig.~\ref{fig:pipeline}: (A)~unified preprocessing method which is device-agnostic, (B)~self-supervised representation learning, and (C)~knowledge distillation with two strategies.

% \begin{figure*}[!t]
%     \centering
%     \fbox{\parbox{0.95\textwidth}{\centering\vspace{2cm}\todo{INSERT PIPELINE OVERVIEW FIGURE --- 3 stages: Preprocessing $\to$ SSL Pretraining $\to$ KD (Strategy A \& B). Show model family from 11.84M to 0.48M.}\vspace{2cm}}}
%     \caption{Overview of the proposed pipeline. Stage~A: heterogeneous EEG signals are aligned to a 62-channel STFT spectrogram representation. Stage~B: an SE-ResNet18 teacher is pretrained via SimCLR on pooled unlabeled data. Stage~C: two KD strategies compress the teacher into compact students---Strategy~A distills pretrained representations into a task-agnostic backbone; Strategy~B distills per-task fine-tuned teachers into task-specialized students.}
%     \label{fig:pipeline}
% \end{figure*}

\vspace{-4mm}
\subsection{Unified Preprocessing}

EEG datasets are recorded using different devices, with varying numbers of channels and sampling rates, making cross-dataset learning challenging. To handle this, we align all recordings to a standard 62-channel layout based on the international 10--10 system~\cite{jurcak200710}. 
Channels present in the source dataset are mapped to their corresponding standard positions, and channels absent in a given dataset are zero-filled rather than interpolated, to avoid introducing synthetic spatial correlations. The SE blocks are expected to learn to down-weight these uninformative channels. Next, as shown in Step 1 of Fig.~\ref{fig:pipeline}, each aligned recording is segmented into non-overlapping 4-second windows, long enough to capture slow delta-band activity (${\sim}$1\,Hz) while remaining short enough for practical processing. We then apply standard preprocessing techniques such as bandpass filtering at 1--50\,Hz (4th-order Butterworth), notch filtering at 50 or 60\,Hz depending on the recording country ($Q{=}30$), baseline correction, and per-channel z-score normalization. The z-normalization step is particularly important in a multi-dataset setting, as it removes amplitude differences caused by varying electrode impedances and hardware gain settings across devices.
Each window is then converted into a time-frequency representation via the Short-Time Fourier Transform. % shown in Eq.~\ref{eq:stft} where $x_c$ is the filtered signal for channel $c$, $w$ is a Hann window of length $L{=}256$ samples, and $H$ is the hop size (25\% of $L$).
% \begin{equation}
%     S_c(f,\,t) = \left|
%         \sum_{n=0}^{L-1} x_c[n + tH]\, w[n]\, e^{-j2\pi fn/L}
%     \right|^2
%     \label{eq:stft}
% \end{equation}
A Hann window is used to minimize spectral leakage between adjacent EEG frequency bands, such as alpha (8--13\,Hz) and beta (13--30\,Hz), whose relative power carries diagnostic meaning~\cite{harris1978windows}. The resulting power spectrogram is cropped to 1--50\,Hz, log-scaled as $\hat{S}_c = \log_{10}(S_c + 10^{-10})$ to compress \mbox{EEG's} characteristically large dynamic range, and resized to $128{\times}128$. The final output per window is a tensor of shape $(62,\,128,\,128)$, one time-frequency image per channel, providing a uniform input format for the model regardless of the originating dataset. Fig.~\ref{fig:preprocess} visualizes example signals before and after preprocessing.

\vspace{-3mm}
\subsection{Self-Supervised Pretraining with SE-ResNet18}

\begin{figure}[t]
    \centering
    \includegraphics[width=1\linewidth]{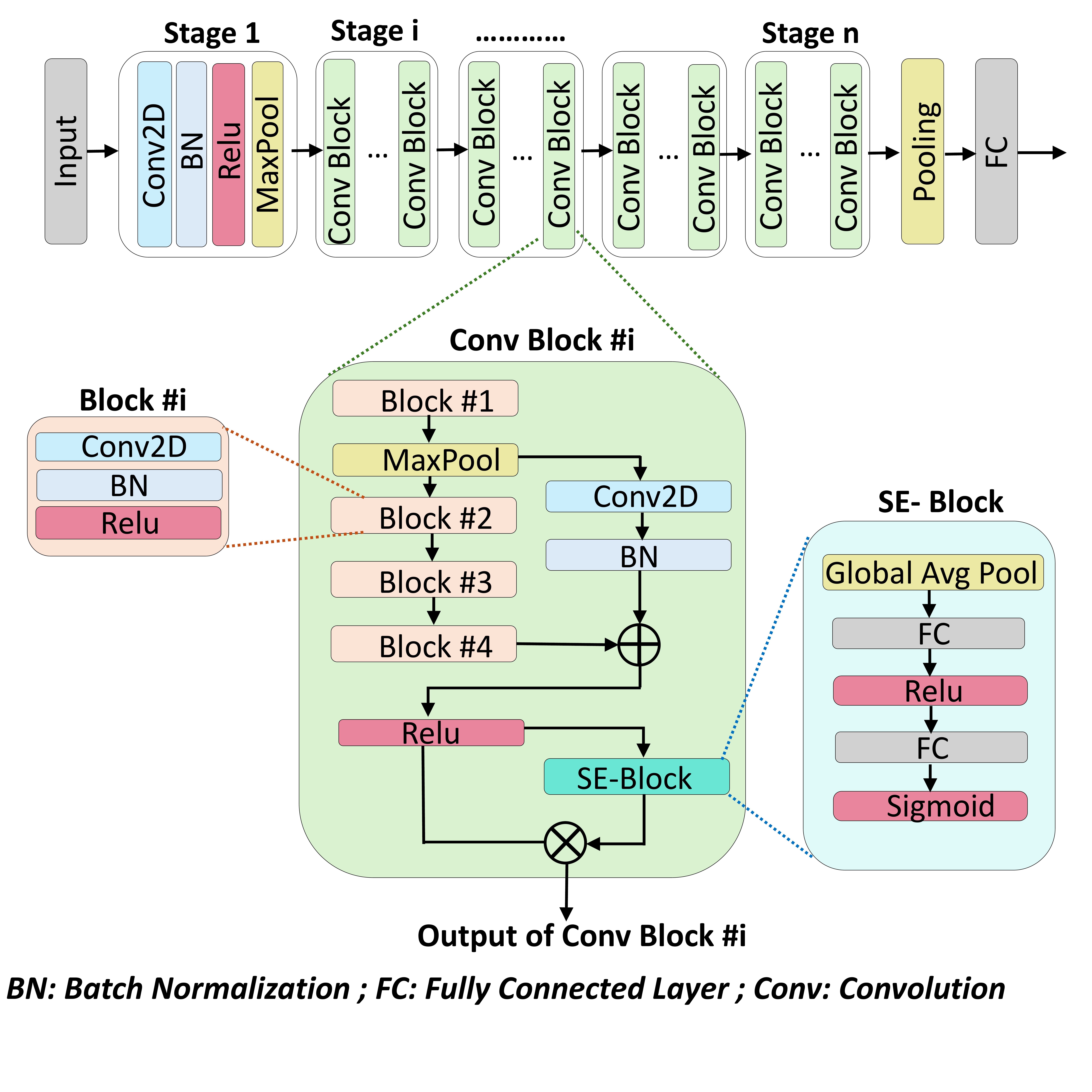}
    % \caption{SE-ResNet encoder architecture. Each residual stage contains stacked convolutional blocks with a squeeze-and-excitation (SE) block that recalibrates channel-wise features via global average pooling and learned gating.}
    \vspace{-11mm}
    \caption{SE-ResNet encoder architecture. SE blocks recalibrate channel-wise features after each residual stage.}
    ~\vspace{-8mm}
    \label{fig:ssl_base}
\end{figure}

The next stage trains a single encoder that generalizes across EEG tasks without relying on task labels. Self-supervised pretraining is well-suited for this goal, as it learns representations from the statistical structure of the data, producing embeddings that capture general neural dynamics transferable across downstream tasks. We adopt SimCLR~\cite{chen2020simclr} for its simplicity and strong performance: it uses a shared encoder, a lightweight projection head, and a contrastive loss on augmented view pairs, without requiring memory banks, momentum encoders, or auxiliary networks. This simplicity is beneficial in our setting for two reasons. First, it reduces hyperparameter sensitivity when training on heterogeneous, multi-dataset EEG data, where batch composition and augmentation statistics vary. Second, the absence of auxiliary components reduces memory overhead for 62-channel spectrogram inputs and simplifies the subsequent distillation stage, since only a single encoder needs to be compressed. We pretrain an SE-ResNet18 encoder on pooled multi-dataset EEG with all task labels removed, enabling the model to learn shared spectro-temporal structure across populations, devices, and cognitive conditions. The backbone is a ResNet18 modified with squeeze-and-excitation (SE) blocks after each of four residual stages (Fig.~\ref{fig:ssl_base}). SE blocks dynamically recalibrate channel-wise feature responses via global average pooling and a learned gating mechanism, allowing the network to emphasize informative channels and suppress noisy or interpolated ones~\cite{hu2018senet}. In our prior conference paper~\cite{chowdhury2025sslse}, we showed that integrating SE blocks with SSL consistently improves EEG classification accuracy while adding a power overhead of ${\leq}$0.4\%. This makes SE attention especially worthwhile in our 62-channel setting, where channel quality is inherently uneven due to interpolation. The encoder maps $(62, 128, 128)$ spectrograms to 512-d embeddings.

% \begin{table}[!t]
%     \centering
%     \caption{Model family. All models retain SE blocks for channel attention. A non-SE ResNet8 ablation quantifies SE overhead separately.}
%     \label{tab:model-summary}
%     \begin{tabular}{lcccc}
%         \toprule
%         \textbf{Model} & \textbf{Params} & \makecell{\textbf{Size}\\\textbf{(MB)}} & \textbf{SE?} & \textbf{Compr.} \\
%         \midrule
%         SE-ResNet18 (T) & 11.84 M & ${\sim}$47 & $\checkmark$ & 1$\times$ \\
%         SE-ResNet8 (S)  & 1.56 M  & ${\sim}$6  & $\checkmark$ & 7.6$\times$ \\
%         SE-ResNet4 (S)  & 0.48 M  & ${\sim}$2  & $\checkmark$ & 24.7$\times$ \\
%         \midrule
%         ResNet8 (ablation) & 1.51 M & ${\sim}$6 & -- & 7.8$\times$ \\
%         \bottomrule
%     \end{tabular}
% \end{table}

For contrastive pretraining, we use spectrograms from all pretraining datasets, discarding labels. Two augmented views of each spectrogram are generated through channel dropout ($p{=}0.1$, simulating electrode detachment), time-bin masking (15\% of bins, $p{=}0.3$, simulating motion artifacts), frequency-bin masking (15\% of bins, $p{=}0.3$), and additive Gaussian noise ($\sigma{=}0.02$,  simulating environmental interference). This pipeline is applied twice per sample, so the two views of the same sample will almost always differ, which is the core mechanism driving contrastive learning. Both views then pass through the shared encoder and a projection MLP. The projection head is discarded after pretraining, and the encoder's 512-d output is used for downstream tasks, following the finding in~\cite{chen2020simclr} that the projection head absorbs task-irrelevant information loss. We use the normalized temperature-scaled cross-entropy (NT-Xent) contrastive loss (Eq.~\ref {eq:ntxent}) to maximize agreement between views of the same spectrogram. 
~\vspace{-1mm}
\begin{equation}
    \mathcal{L}_{i,j} = -\log \frac{\exp(\mathrm{sim}(\mathbf{z}_i, \mathbf{z}_j) / \tau)}{\sum_{k=1}^{2N} \mathbf{1}_{[k \neq i]} \exp(\mathrm{sim}(\mathbf{z}_i, \mathbf{z}_k) / \tau)}
    \label{eq:ntxent}
\end{equation}

Here, $\mathbf{z}_i$ and $\mathbf{z}_j$ denote the projected embeddings of two augmented views, $\mathrm{sim}(\cdot,\cdot)$ is cosine similarity, and $\tau$ is a temperature parameter controlling the sharpness of the similarity distribution. The denominator sums over all $2N$ embeddings in the minibatch, excluding $\mathbf{z}_i$. The temperature $\tau$ plays a critical role in our multi-dataset setting: lower values sharpen the distribution, encouraging finer-grained distinctions between spectrograms. This helps separate semantically similar yet diagnostically distinct patterns (e.g., healthy alpha activity vs.\ pathological slowing) that might otherwise collapse in the embedding space. The use of cosine similarity further improves robustness to residual amplitude variations across datasets, which may persist after normalization. Additionally, since all samples in the minibatch act as negatives regardless of their dataset of origin, the loss implicitly enforces discrimination across recording conditions, devices, and populations. This cross-dataset contrastive pressure enables the encoder to learn transferable representations that generalize to unseen tasks and acquisition settings.
~\vspace{-3mm}

% The pretraining pool and evaluation datasets are entirely disjoint (Section~\ref{sec:setup}).

\begin{figure}[t]
    \centering
    \includegraphics[width=1\linewidth]{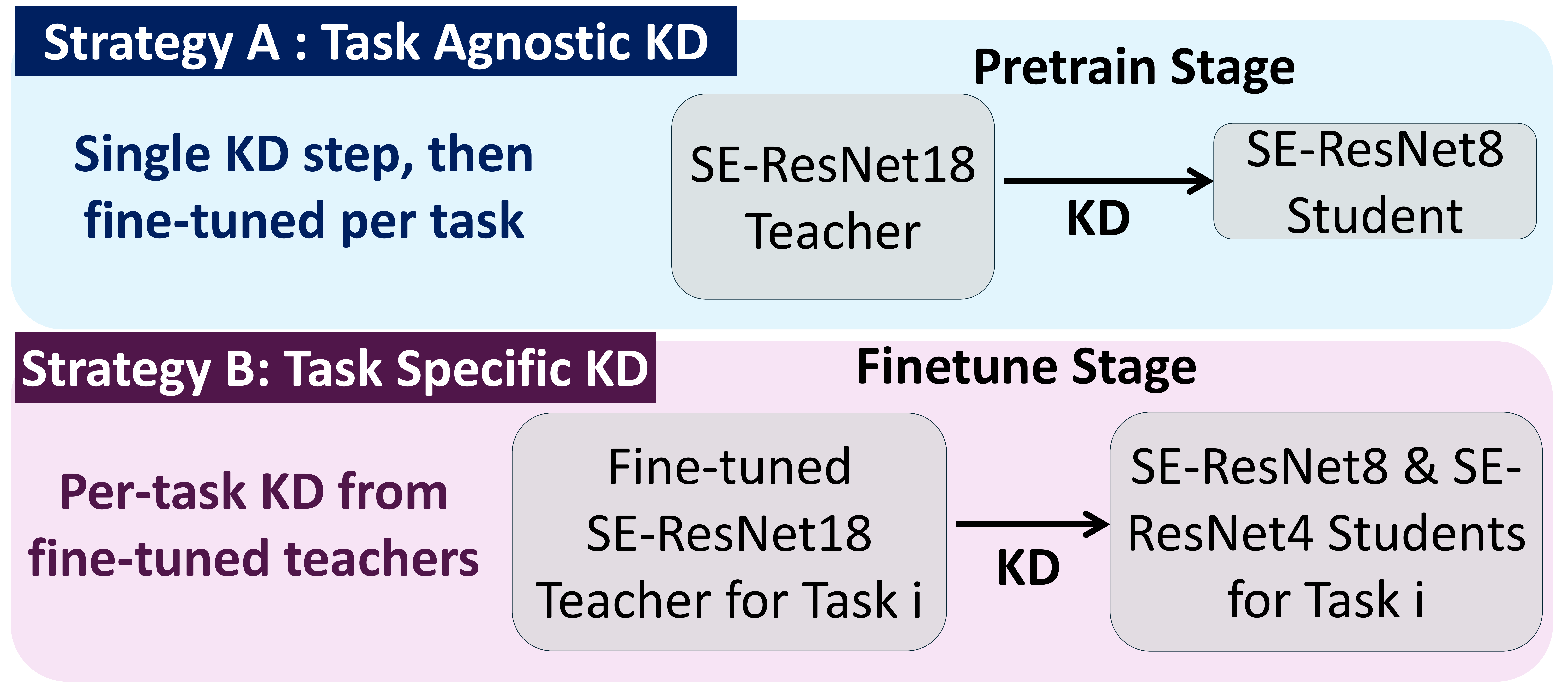}
    \caption{Two KD strategies: Strategy~A distills once from the SSL-pretrained teacher; Strategy~B distills separately from each fine-tuned teacher.}
    \vspace{-5mm}
    \label{fig:kd}
\end{figure}

\vspace{-2mm}
\subsection{Knowledge Distillation (KD)}
\label{sub_sec:kd}

The final stage in our pipeline is knowledge distillation. We do this because the pretrained model is too large for efficient deployment at 11.84\,M parameters (${\sim}$44\,MB). To this end, we compress the teacher into smaller students, SE-ResNet8 (1.56\,M) and SE-ResNet4 (0.48\,M), produced through different distillation strategies. Both students retain SE blocks, given the favorable accuracy-to-overhead tradeoff established in~\cite{chowdhury2025sslse}. We investigate two distillation strategies as shown in Fig.~\ref{fig:kd}. 
Strategy~A compresses the SSL-pretrained teacher through one round of task-agnostic distillation into a shared backbone, which is then fine-tuned per downstream task. Strategy~B instead first fine-tunes the teacher per task, then distills each fine-tuned teacher into a separate student, requiring per-task KD but allowing the student to benefit from task-specific knowledge. Together, these strategies span different points on the accuracy--efficiency tradeoff, characterized across hardware tiers in Section~\ref{sec:results}.

\textbf{Strategy~A (Task-agnostic).} We distill the SSL-pretrained SE-ResNet18 into an SE-ResNet8 student (256-d embeddings) on the same unlabeled pretraining data. Since the student and teacher have different embedding dimensions, the student is augmented with a linear projection (256-d~$\to$~512-d) during distillation; this layer is discarded afterwards. The loss combines NT-Xent and MSE, $\mathcal{L}_{\mathrm{A}} = \mathcal{L}_{\mathrm{NT\text{-}Xent}}(z_S, z_S') + \lambda\,\mathrm{MSE}(p_S, f_T(x))$, where $z_S$ and $z_S'$ are the student's projected embeddings of two augmented views, $p_S$ is the student's projection into teacher space, and $f_T(x)$ is the frozen teacher's embedding. The NT-Xent term preserves view-invariant learning, while the MSE term anchors the student to the teacher's embedding space; both terms are computed over both augmented views and averaged. The resulting backbone serves as a shared initialization for downstream fine-tuning.

\textbf{Strategy~B (Task-specific).} The teacher is first fine-tuned per downstream task, then each task-specific teacher (frozen) distills into an SE-ResNet8 and an SE-ResNet4 student using the standard Hinton KD loss~\cite{hinton2015distilling}, $\mathcal{L}_{\mathrm{B}} = \alpha \cdot \mathcal{L}_{\mathrm{CE}} + (1 - \alpha) \cdot T^{2} \cdot \mathrm{KL}(\sigma(z_S/T) \| \sigma(z_T/T))$, where $T$ controls distribution softening and $\alpha$ balances the two terms.

Since the teacher has been fine-tuned on task labels, the student benefits from both soft distributional knowledge and direct supervision, helping it learn more calibrated decision boundaries than hard-label training alone. However, Strategy~B relies on the quality of the teacher's soft labels, so it is most effective when the downstream dataset is sufficiently large to yield reliable class boundaries. When labeled data is scarce, the teacher's outputs may be noisy, in which case Strategy~A's task-agnostic initialization can serve as a better starting point.
~

\vspace{-2mm}
\section{Experimental Setup}
\label{sec:setup}
Having described the framework, we now present the datasets, training protocol, and hardware profiling setup used to evaluate the proposed approach.

\noindent\textbf{Datasets}: For this work, we used open source EEG datasets for three common BCI applications, namely emotion recognition, motor imagery and abnormality detection. Table~\ref{tab:datasets} lists the dataset details. The pretraining pool and evaluation sets share no datasets in common. The pretraining pool spans five datasets with 3--64 channels and 128--512\,Hz sampling rates. Six separate datasets, two from each application category, are used for downstream evaluation.

\begin{table}[!t]
    \centering
    \caption{Overview of datasets used in this work.}
    ~\vspace{-3mm}
    \label{tab:datasets}
    \begin{tabular}{llccc}
        \toprule
        \textbf{Category} & \textbf{Dataset} & \textbf{\# Classes} & \textbf{\# Channels} & \textbf{$f_s$ (Hz)} \\
        \midrule
        \multicolumn{5}{l}{\textit{SSL Pretraining Pool (labels discarded)}} \\
        \midrule
        Emotion     & SEED-IV      & 4 & 62 & 200  \\
        Emotion     & DEAP         & 2 & 32 & 128  \\
        Motor Im.   & PhysioNet MI & 4 & 64 & 160  \\
        Abnormality & TUH-Sz       & 2 & 19--128 & 250--512 \\
        Abnormality & CHB-MIT      & 2 & 23 & 256  \\
        \midrule
        \multicolumn{5}{l}{\textit{Evaluation Datasets (fine-tuned with labels)}} \\
        \midrule
        Abnormality & TUAB         & 2 & 21 & var \\
        Abnormality & SIENA        & 2 & 29 & 512   \\
        Motor Im.   & BCI-IV-2a    & 4 & 22 & 250   \\
        Motor Im.   & BCI-IV-2b    & 2 & 3  & 250   \\
        Emotion     & SEED         & 3 & 62 & 200   \\
        Emotion     & EmoEEG       & 7 & 28 & 250   \\
        \bottomrule
    \end{tabular}
    ~\vspace{-2mm}
\end{table}

%\subsection{Training Details}

\noindent\textbf{SSL pretraining}: The teacher model is pretrained for 50 epochs on 4 NVIDIA L40S GPUs with a total batch size of 1024. We adopt SGD with a base learning rate of $3 \times 10^{-2}$, scaled from a reference batch size of 256. This choice leverages the implicit gradient noise of SGD, which helps mitigate representational collapse in contrastive learning. Training employs a 2-epoch linear warmup followed by cosine annealing. We set the weight decay to $10^{-4}$ and the momentum to 0.9, and use FP16 mixed-precision to improve training efficiency. The projection MLP maps the 512-d encoder output to a 128-d embedding space, following SimCLR, where the projection head helps absorb task-irrelevant information. During fine-tuning, the projection head is discarded, and only the encoder backbone is retained.

%Please add student model details.

% \noindent\textbf{Fine-tuning}: The SSL-pretrained teacher is fine-tuned end-to-end on each downstream dataset for 50 epochs. We adopt a three-tier layer-wise learning-rate schedule, using \(10^{-5}\) for the lower layers, \(10^{-4}\) for the upper convolutional layers, and \(10^{-3}\) for the classification head, so as to preserve broadly transferable low-level representations while allowing greater task-specific adaptation in higher layers~\cite{howard2018ulmfit}. To address class imbalance, we use class-weighted cross-entropy with inverse-frequency weighting. Early stopping with a patience of 10 epochs is applied to reduce overfitting, which is especially important for smaller datasets such as EmoEEG. 

\noindent\textbf{Fine-tuning}: The SSL-pretrained teacher is fine-tuned end-to-end on each downstream dataset for 50 epochs with a three-tier layer-wise learning-rate schedule: \(10^{-5}\) for the lower layers, \(10^{-4}\) for the upper convolutional layers, and \(10^{-3}\) for the classification head, preserving transferable low-level representations while allowing task-specific adaptation in higher layers~\cite{howard2018ulmfit}. A linear classification head maps the encoder output (512-d for the teacher, 256-d for SE-ResNet8, 128-d for SE-ResNet4) to the number of downstream classes. We use class-weighted cross-entropy with inverse-frequency weighting to address class imbalance, and early stopping with patience of 10 epochs to reduce overfitting. The student models follow the same training protocol in both distillation strategies.

\noindent\textbf{Baseline models}: We compare against two groups of models. \emph{Internal baselines} include our teacher and student models trained under different settings, such as SSL and KD-based fine-tuning. \emph{External baselines} include both EEG FMs and task-specific architectures drawn from prior work. The exact set of external baselines varies by dataset and includes state-of-the-art EEG FMs such as LaBraM~\cite{jiang2024labram}, BIOT~\cite{yang2023biot}, and CBraMod~\cite{wang2025cbramod}, as well as task-specific methods like EEGNet~\cite{lawhern2018eegnet} and EEGConformer~\cite{song2023conformer} when available. External numbers are taken from the corresponding published papers.  

% EEGNet~\cite{lawhern2018eegnet}, EEGConformer~\cite{song2023conformer},CSBrain~\cite{zhou2025csbrain},REVE~\cite{ouahidi2025reve}

\noindent\textbf{Inference Profiling}: We profile inference on three platforms using batch size 1 to reflect single-sample, real-time inference. The evaluated hardware consists of: (1)~\textbf{Server-grade GPU}: an NVIDIA L4, with power sampled via NVML (\texttt{pynvml}); (2)~\textbf{CPU}: an Intel i7-8700 (3.20\,GHz, 65\,W TDP), for which power is estimated using a utilization-scaled TDP proxy derived from \texttt{psutil}; and \textcolor{black}{(3)~\textbf{Edge Device}: an NVIDIA Jetson Orin Nano edge platform (Ampere architecture GPU with ARM Cortex-A78AE CPU cores, configurable low-power modes in the 7–15 W range), with power monitored via \texttt{jtop} and onboard INA telemetry.}~\vspace{-3mm}
% Each experiment includes 50 warmup inferences followed by 200 timed runs for latency measurement and 500 runs for power estimation. We report latency, power and energy per inference.

% for which power is estimated using a utilization-scaled TDP proxy derived from \texttt{psutil} (this proxy may overestimate dynamic power, as TDP reflects worst-case thermal design rather than typical operating power)

%% ====================================================================
%%  V. RESULTS
%% ====================================================================
~\vspace{-4mm}
\section{Results}
\label{sec:results}

With the experimental setup established, we now present results across all six downstream benchmarks, both KD strategies, and three deployment-oriented hardware tiers. We begin by comparing our models with state-of-the-art FMs, then examine how well performance is retained under compression, and finally assess whether the resulting compact models deliver meaningful efficiency gains on real hardware.

\begin{table}[t]
    \centering
    \footnotesize
    \caption{SOTA comparison on abnormality detection.}
    \label{tab:sota-abnormality}
    \begin{tabular}{llcc}
        \toprule
        \textbf{Dataset} & \textbf{Method} & \textbf{Params} & \textbf{Acc (\%)} \\
        \midrule
        \multirow[t]{21}{*}[-0.7ex]{TUAB}
        %% Task-specific (standard order) %%
        & EEGNet~\cite{lawhern2018eegnet}\textsuperscript{a}           & 0.003 M  & 76.42 \\
        & EEGConformer~\cite{zhou2025csbrain,song2023conformer}\textsuperscript{a}     & 0.55 M   & 77.58 \\
        & SPaRCNet~\cite{jing2023development}\textsuperscript{a}       & 0.79 M   & 78.96 \\
        & ContraWR~\cite{yang2023self}\textsuperscript{a}              & 1.6 M    & 77.46 \\
        & CNN-Transformer~\cite{peh2022transformer}\textsuperscript{a} & 3.2 M    & 77.77 \\
        & FFCL~\cite{li2022motor}\textsuperscript{a}                   & 2.4 M    & 78.48 \\
        % & ST-Transformer\textsuperscript{a}                            & 3.5 M    & 79.66 \\
        %% FMs (ascending param count) %%
        & BIOT~\cite{yang2023biot}                                     & 3.2 M    & 79.59 \\
       
        & CBraMod~\cite{wang2025cbramod}                               & 4.0 M    & 82.89 \\
         & CSBrain~\cite{zhou2025csbrain} 
        & 4.9 M & 81.72 \\
        
        & LaBraM-Base~\cite{jiang2024labram}                           & 5.8 M    & 81.40 \\
        & LaBraM-Large~\cite{jiang2024labram}                          & 46 M     & 82.26 \\
        & REVE~\cite{ouahidi2025reve}                                  & 69 M     & 83.15 \\
        & NeuroLM-B~\cite{wei2025neurolm}                              & 254 M    & 78.26 \\
        & LaBraM-Huge~\cite{jiang2024labram}                           & 369 M    & 82.58 \\
        & NeuroLM-L~\cite{wei2025neurolm}                              & 508 M    & 78.68 \\
        & NeuroLM-XL~\cite{wei2025neurolm}                             & 1{,}696 M & 79.69 \\
        \cmidrule(lr){2-4}
        &  (BRIDGE-EEG) Teacher                                                 & 11.84 M  & 86.94 \\
        &  (BRIDGE-EEG) SE-ResNet8 (A)                                          & 1.56 M   & 89.75 \\
        &  (BRIDGE-EEG) SE-ResNet8 (B)                                          & 1.56 M   & \textbf{90.35} \\
        &  (BRIDGE-EEG) SE-ResNet4 (B)                                          & 0.48 M   & 88.00 \\
       
        \midrule
        \multirow[t]{22}{*}[-0.7ex]{SIENA}
        %% Task-specific (standard order, commented) %%
         % & EEGNet~\cite{lawhern2018eegnet}\textsuperscript{a}           & 0.003 M  & 74.87 \\
        % & EEGConformer~\cite{song2023conformer}\textsuperscript{a}     & 0.55 M   & 75.56 \\
        % & SPaRCNet~\cite{jing2023development}\textsuperscript{a}       & 0.79 M   & 65.72 \\
        % & ContraWR~\cite{yang2023self}\textsuperscript{a}              & 1.6 M    & 65.46 \\
        % & CNN-Transformer~\cite{peh2022transformer}\textsuperscript{a} & 3.2 M    & 69.82 \\
        % & FFCL~\cite{li2022motor}\textsuperscript{a}                   & 2.4 M    & 66.16 \\
        % & ST-Transformer\textsuperscript{a}                            & 3.5 M    & 75.27 \\
        %% FMs (ascending param count) %%
        & BIOT~\cite{yang2023biot}                                     & 3.2 M    & 73.52 \\
        & CBraMod~\cite{wang2025cbramod}                               & 4.0 M    & 73.17 \\
        & CSBrain~\cite{zhou2025csbrain}                               & 4.9 M    & 76.62 \\
        & LaBraM-Base~\cite{jiang2024labram}                           & 5.8 M    & 70.82 \\
        %% Dataset-specific task-specific %%
        & Phase Amplitude Coupling + Rs~\cite{jiang2023epileptic}\textsuperscript{a}           & --        & 85.71 \\
        & GoogLeNet~\cite{wang2023automatic}\textsuperscript{a}         & --        & 96.42 \\
        & Bi-LSTM + 1D-CNN~\cite{yang2023patient}\textsuperscript{a}   & --        & \textbf{99.70} \\
        \cmidrule(lr){2-4}
        &  (BRIDGE-EEG) Teacher                                                 & 11.84 M  & 98.40 \\
        &  (BRIDGE-EEG) SE-ResNet8 (A)                                          & 1.56 M   & 99.50 \\
        &  (BRIDGE-EEG) SE-ResNet8 (B)                                          & 1.56 M   & 99.54 \\
        &  (BRIDGE-EEG) SE-ResNet4 (B)                                          & 0.48 M   & 99.08 \\
        
        \bottomrule
        \multicolumn{4}{l}{\footnotesize (A) = KD on pretrained. (B) = KD on fine-tuned. \textsuperscript{a} Task-specific model.
        ~\vspace{-5mm}} \\
    \end{tabular}
    
\end{table}

\begin{table}[t]
    \centering
    \footnotesize
    \caption{SOTA comparison on emotion detection.}
    \label{tab:sota-emotion}
    \begin{tabular}{llcc}
        \toprule
        \textbf{Dataset} & \textbf{Method} & \textbf{Params} & \textbf{Acc (\%)} \\
        \midrule
        \multirow[t]{14}{*}[-0.7ex]{SEED}
        %% Task-specific (standard order) %%
        % & EEGNet~\cite{lawhern2018eegnet}\textsuperscript{a}           & 0.003 M  & -- \\
        % & EEGConformer~\cite{song2023conformer}\textsuperscript{a}     & 0.55 M   & -- \\
        & SPaRCNet~\cite{jing2023development}\textsuperscript{a}       & 0.79 M   & 55.96 \\
        & ContraWR~\cite{yang2023self}\textsuperscript{a}              & 1.6 M    & 61.06 \\
        & CNN-Transformer~\cite{peh2022transformer}\textsuperscript{a} & 3.2 M    & 61.61 \\
        & FFCL~\cite{li2022motor}\textsuperscript{a}                   & 2.4 M    & 58.08 \\
        & ST-Transformer\textsuperscript{a}                            & 3.5 M    & 54.79 \\
        %% FMs (ascending param count) %%
        & BIOT~\cite{yang2023biot}                                     & 3.2 M    & 70.97 \\
        & LaBraM-Base~\cite{jiang2024labram}                           & 5.8 M    & 73.18 \\
        & NeuroLM-B~\cite{wei2025neurolm}                              & 254 M    & 55.54 \\
        & NeuroLM-L~\cite{wei2025neurolm}                              & 508 M    & 60.06 \\
        & NeuroLM-XL~\cite{wei2025neurolm}                             & 1{,}696 M & 60.34 \\
        \cmidrule(lr){2-4}
         &  (BRIDGE-EEG) Teacher                                                 & 11.84 M  & \textbf{87.68} \\
          &  (BRIDGE-EEG) SE-ResNet8 (A)                                          & 1.56 M   & 78.10 \\
        &  (BRIDGE-EEG) SE-ResNet8 (B)                                          & 1.56 M   & 80.00 \\
        &  (BRIDGE-EEG) SE-ResNet4 (B)                                          & 0.48 M   & 70.00 \\
       
        \midrule
        \multirow[t]{4}{*}[-0.7ex]{EmoEEG}
        &  (BRIDGE-EEG) Teacher                                                 & 11.84 M  & 75.89 \\
        &  (BRIDGE-EEG) SE-ResNet8 (A)                                          & 1.56 M   & \textbf{77.32} \\
        &  (BRIDGE-EEG) SE-ResNet8 (B)                                          & 1.56 M   & 60.29 \\
        &  (BRIDGE-EEG) SE-ResNet4 (B)                                          & 0.48 M   & 35.10 \\
        \bottomrule
        \multicolumn{4}{l}{\footnotesize (A) = KD on pretrained. (B) = KD on fine-tuned. \textsuperscript{a} Task-specific model.} \\
    \end{tabular}
    ~\vspace{-2mm}
\end{table}

\begin{table}[t]
    \centering
    \footnotesize
    \caption{SOTA comparison on motor imagery.}
    \label{tab:sota-mi}
    \begin{tabular}{llcc}
        \toprule
        \textbf{Dataset} & \textbf{Method} & \textbf{Params} & \textbf{Acc (\%)} \\
        \midrule
        \multirow[t]{17}{*}[-0.7ex]{BCI-IV-2a}
        %% Task-specific (standard order) %%
        % & EEGNet~\cite{lawhern2018eegnet}\textsuperscript{a}           & 0.003 M    & 51.30 \\
        % & EEGConformer~\cite{song2023conformer}\textsuperscript{a}     & 0.55 M    & -- \\
        % & SPaRCNet~\cite{jing2023development}\textsuperscript{a}       & 0.79 M    & -- \\
        % & ContraWR~\cite{yang2023self}\textsuperscript{a}              & 1.6 M     & -- \\
        & CNN-Transformer~\cite{peh2022transformer}\textsuperscript{a} & 3.2 M      & 47.70 \\
        % & FFCL~\cite{li2022motor}\textsuperscript{a}                   & 2.4 M     & -- \\
        % & ST-Transformer\textsuperscript{a}                            & 3.5 M     & -- \\
        %% FMs (ascending param count) %%
        & CBraMod~\cite{wang2025cbramod}                               & 4.0 M     & 51.38 \\
        & EEGPT~\cite{wang2024eegpt}                                    & 4.7--25 M & 58.46 \\
        & LaBraM~\cite{jiang2024labram}                                & 5.8 M     & 56.13 \\
        & REVE~\cite{ouahidi2025reve}                                  & 69 M      & \textbf{63.96} \\
        & NeuroGPT~\cite{cui2024neurogpt}                              & 79.53 M   & 58.60 \\
        & BENDR~\cite{kostas2021bendr,wang2024eegpt}                   & $>$1\,B    & 48.99 \\
        \cmidrule(lr){2-4}
        &  (BRIDGE-EEG) Teacher                                                 & 11.84 M   & 40.71 \\
         &  (BRIDGE-EEG) SE-ResNet8 (A)                                          & 1.56 M    & 38.51 \\
        &  (BRIDGE-EEG) SE-ResNet8 (B)                                          & 1.56 M    & 38.66 \\
        &  (BRIDGE-EEG) SE-ResNet4 (B)                                          & 0.48 M    & 35.13 \\
       
        \midrule
        \multirow[t]{8}{*}[-0.7ex]{BCI-IV-2b}
        %% FMs (ascending param count) %%
        & BIOT~\cite{yang2023biot}                                     & 3.2 M    & 64.09 \\
        & LaBraM~\cite{jiang2024labram}                                & 5.8 M    & 68.51 \\
        & EEGPT~\cite{wang2024eegpt}                                    & 25 M     & \textbf{72.12} \\
        & BENDR~\cite{kostas2021bendr,wang2024eegpt}                   & $>$1\,B   & 70.67 \\
        \cmidrule(lr){2-4}
        &  (BRIDGE-EEG) Teacher                                                 & 11.84 M  & 70.86 \\
        &  (BRIDGE-EEG) SE-ResNet8 (A)                                          & 1.56 M   & 69.00 \\
        &  (BRIDGE-EEG) SE-ResNet8 (B)                                          & 1.56 M   & 68.05 \\
         &  (BRIDGE-EEG) SE-ResNet4 (B)                                          & 0.48 M   & 66.89 \\
        \bottomrule
        \multicolumn{4}{l}{\footnotesize (A) = KD on pretrained. (B) = KD on fine-tuned. \textsuperscript{a} Task-specific model.} \\
    \end{tabular}
    ~\vspace{-5mm}
\end{table}

\vspace{-2mm}
\subsection{EEG-FM like accuracy}
\label{subsec:sota}

Tables~\ref{tab:sota-abnormality},~\ref{tab:sota-emotion}, and~\ref{tab:sota-mi} present the comparison of BRIDGE-EEG with state-of-the-art EEG FMs and, where FM baselines are unavailable, with task-specific architectures. 
% It is worth noting that the compared FMs range from 3.2 M to 1{,}696 M parameters, whereas our distilled SE-ResNet8 and SE-ResNet4 students contain only 1.56 M and 0.48 M parameters, respectively.

% \textbf{Abnormality detection.} The results of abnormality detection are presented in Table~\ref{tab:sota-abnormality}. On TUAB, the Strategy~B SE-ResNet8 student achieves 90.35\% accuracy with 1.56M parameters, which is comparable to the FMs including REVE (83.15\%, 69M) and LaBraM-Huge (82.58\%, 369M), while using approximately 44$\times$ and 237$\times$ fewer parameters, respectively. SE-ResNet4, with only 0.48 M parameters, achieves 88.00\%, which is still above all benchmarked FMs. On SIENA, the SE-ResNet8~(B) student reaches 99.54\%, which is within 0.16 percentage points of the best reported task-specific result of Bi-LSTM + CWT (99.70\%).

% Both SE-ResNet8 students outperform the teacher (87\%) on this binary task, which may reflect a regularizing effect of distillation or an implicit capacity benefit of the smaller architecture.

\textbf{Abnormality detection.} The results of abnormality detection are presented in Table~\ref{tab:sota-abnormality}. On TUAB, the Strategy~B SE-ResNet8 student achieves 90\% accuracy with 1.56\,M parameters, which is above all benchmarked EEG FMs, including REVE (83\%, 69\,M) and LaBraM-Huge (83\%, 369\,M), while using approximately 44$\times$ and 237$\times$ fewer parameters, respectively. Both SE-ResNet8 students outperform the teacher (87\%) on this binary task, consistent with the known regularizing effect of knowledge distillation observed in prior work~\cite{hinton2015distilling}. SE-ResNet4, with only 0.48\,M parameters, achieves 88\%, still above all benchmarked FMs. The task-specific baselines (e.g., EEGNet, EEGConformer, SPaRCNet), which are trained from scratch without pretraining, all fall below 79\% on TUAB. On SIENA, the SE-ResNet8~(B) student reaches 99\%, within 0.16 percentage points of the best reported task-specific result (Bi-LSTM + CWT, 99\%) and above all FM baselines. The strong abnormality-detection results may reflect both the pretraining corpus composition (Table~\ref{tab:datasets}), which includes two abnormality datasets (TUH-Sz, CHB-MIT), and the fact that dataset-specific spectral patterns are likely well preserved in the unified 62-channel representation.

% Both SE-ResNet8 students perform better than the teacher (87\%), suggesting that distillation may provide a regularizing effect on this binary task.
% \textbf{Emotion recognition.} A similar pattern appears on emotion recognition (Table~\ref{tab:sota-emotion}). On SEED, the teacher achieves 87.68\%, and the distilled SE-ResNet8 students score 80.00\% (Strategy~B) and 78.10\% (Strategy~A). Both are above LaBraM-Base (73.18\%, 5.8M) and BIOT (70.97\%, 3.2M), as well as NeuroLM-XL (60.34\%, 1{,}696M). On EmoEEG, Strategy~A (77.32\%) scores higher than both Strategy~B (60.29\%) and the teacher (75.89\%). We attribute this to the limited scale of EmoEEG, which contains recordings from only eight subjects. Under such data-scarce conditions, representation-level transfer (Strategy~A) appears to act as an implicit regularizer. In contrast, task-level distillation (Strategy~B) tends to inherit the fine-tuned teacher's task-specific biases, leading to poorer generalization.

\textbf{Emotion recognition.} Emotion recognition results are presented in Table~\ref{tab:sota-emotion}. On SEED, the teacher achieves ${\sim}$88\%, and the distilled SE-ResNet8 students score 80\% (Strategy~B) and 78\% (Strategy~A). While both show a moderate drop from the teacher, they remain above all benchmarked FMs, including LaBraM-Base (73\%, 6\,M) and NeuroLM-XL (60\%, 1{,}696\,M). Task-specific baselines on SEED such as FFCL and CNN-Transformer remain below 62\%, well behind both FMs and our compressed students. On EmoEEG, Strategy~A (77\%) scores higher than both Strategy~B (60\%) and the teacher (76\%), while SE-ResNet4 drops to 35\% on this 7-class task. We attribute the advantage of Strategy~A to the limited scale of EmoEEG (eight subjects). Under such data-scarce conditions, representation-level transfer appears to act as an implicit regularizer, whereas task-specific distillation (Strategy~B) tends to inherit the fine-tuned teacher's biases, and the smallest student likely lacks sufficient capacity for fine-grained emotion discrimination. The fact that Strategy~B still outperforms Strategy~A on SEED, despite SEED also being a relatively small dataset (15 subjects), suggests that dataset size alone does not explain the reversal. Rather, EmoEEG's 7-class structure combined with only eight subjects yields far fewer samples per class, likely making the fine-tuned teacher's soft labels unreliable and favoring the task-agnostic approach. 

% \begin{figure*}[t]
%     \centering
%     \includegraphics[width=0.8\linewidth]{Figs/Fig_KD_res2_1.pdf}
%     ~\vspace{-2mm}
%     \caption{Accuracy retention under compression across all six benchmarks. Students match or exceed the teacher on binary clinical tasks, maintain moderate performance in emotion recognition, and show the largest declines in motor imagery.}
%     ~\vspace{-8mm}
%     \label{fig:kd-results}
    
% \end{figure*}

\begin{figure}[t]
    \centering
    \includegraphics[width=0.85\linewidth]{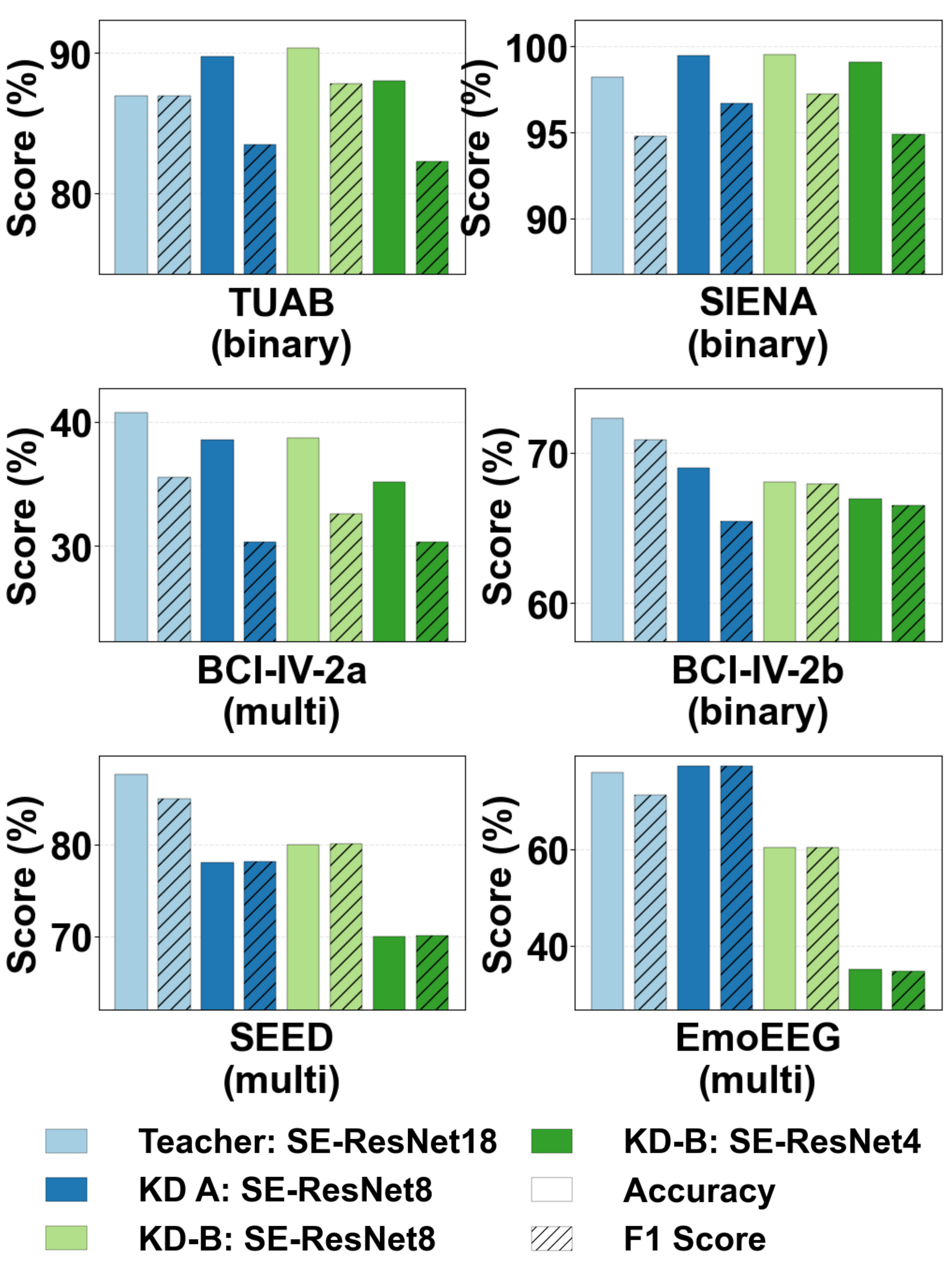}
    ~\vspace{-3mm}
    \caption{Accuracy retention under compression across all six benchmarks. Students match or exceed the teacher on binary clinical tasks, maintain moderate performance in emotion recognition, and show the largest declines in motor imagery.}
    ~\vspace{-8mm}
    \label{fig:kd-results}
    
\end{figure}

\textbf{Motor imagery.} The results of motor imagery classification are presented in Table~\ref{tab:sota-mi}. Although some dataset-specific evaluations, such as~\cite{song2023conformer}, have fewer parameters, they use per-subject evaluations(78.66\% accuracy on BCI-IV-2a), but re-evaluation under generalized protocols yields $\sim$47\%~\cite{zhou2025csbrain}. Under generalized conditions on BCI-IV-2a, our best model, SE-ResNet8~(B), achieves 39\% (chance level: 25\%), which is below several benchmarked FMs, such as REVE (64\%, 69\,M) and NeuroGPT (59\%, 80\,M). We hypothesize that this gap arises from two factors. First, the pretraining corpus includes only a single motor imagery dataset (PhysioNet MI), whereas the abnormality and emotion categories each benefit from two (Table~\ref{tab:datasets}); and secondly, MI datasets use far fewer channels than our 62-channel representation (22 for BCI-IV-2a, 3 for BCI-IV-2b), so the majority of input channels are zero-filled, which may hinder the model from learning the fine-grained spatial patterns that MI classification requires. The gap narrows on BCI-IV-2b, where our teacher achieves 71\%, comparable to EEGPT (72\%, 25\,M), and the SE-ResNet8~(A) student reaches 69\% with 1.56\,M parameters, likely because the simpler binary task places less demand on spatial resolution. Notably, across both MI datasets the students remain close to the teacher (38\% vs.\ 41\% on BCI-IV-2a; 69\% vs.\ 71\% on BCI-IV-2b), indicating that the distillation pipeline preserves the teacher's learned representations even when the teacher itself underperforms. This suggests that the bottleneck lies in the pretrained representation rather than in the compression, motivating future work to expand MI pretraining diversity and to explore spatial-resolution-preserving input representations. 
% Although compact task-specific architectures such as EEG Conformer~\cite{song2023eegconformer} can achieve competitive per-task accuracy (e.g., 78.66% on BCI-IV-2a under per-subject evaluation), independent re-evaluation under cross-dataset protocols yields substantially lower accuracy ($\sim$47% in~\cite{csbrain}). Our students instead derive from a shared pretrained backbone that generalizes across tasks and recording setups without per-user retraining, and Tables~\ref{tab:sota-abnormal}--\ref{tab:sota-mi} report only numbers obtained under such generalized protocols.

Overall, these results suggest that structured knowledge distillation can yield edge-device-scale models that perform comparably to, and in some cases above, FMs with 10--1{,}000$\times$ more parameters on abnormality detection and emotion recognition tasks. On motor imagery, a gap remains, likely due to limited pretraining diversity for this task category. These observations point to a practical accuracy-efficiency trade-off for deployment on resource-constrained hardware.

\vspace{-3mm}
\subsection{Accuracy Retention Under Compression}
\label{subsec:compression}
Having compared our models with existing FMs and task-specific baselines, we now examine how much performance is retained after compression, and whether accuracy alone is sufficient to assess compression quality. Fig.~\ref{fig:kd-results} summarizes results across all six benchmarks. We report macro-averaged F1 to assess per-class balance. The 1.56\,M-parameter SE-ResNet8 reduces model size by $7.6\times$ while preserving useful performance across all three task categories, and the 0.48\,M SE-ResNet4 remains competitive on binary clinical tasks despite much stronger compression.

F1 scores reveal patterns that accuracy alone does not. On TUAB, the students achieve higher accuracy than the teacher (90\% vs.\ 87\%) but lower F1 (0.83 vs.\ 0.87), suggesting the compressed models favor the majority class more heavily, an important consideration for clinical deployment where missing abnormal recordings is costly. On SEED, Strategy~A shows much closer accuracy with F1 agreement (78\% accuracy, 0.77 F1) than Strategy~B (80\% accuracy, 0.60 F1), indicating that Strategy~A produces more balanced predictions across emotion classes even when its accuracy is slightly lower. On EmoEEG, macro F1 exceeds accuracy for the smallest student, reflecting the sensitivity of macro averaging to per-class fluctuations on this small 7-class dataset. Motor imagery remains the most difficult category, with students below the teacher in both metrics, though the gap is smaller than the $7.6\times$ compression might suggest. Overall, compression is most successful on binary clinical tasks, remains viable on emotion recognition, and is less effective on motor imagery. The F1 analysis shows that accuracy alone can overstate the success of compression on imbalanced tasks, and that Strategy~A tends to yield more class-balanced predictions than Strategy~B.

\begin{table*}[t]
    \centering
    \caption{Inference efficiency at batch size 1. Dynamic energy = (avg. power $-$ idle power) $\times$ latency.}
    \label{tab:deployment}
    \renewcommand{\arraystretch}{0.85}
    \resizebox{\textwidth}{!}{%
    \begin{tabular}{lcc ccc ccc ccc}
        \toprule
        & & & \multicolumn{3}{c}{\textbf{GPU (NVIDIA L4)}} & \multicolumn{3}{c}{\textbf{CPU (i7-8700)}} & \multicolumn{3}{c}{\textbf{NVIDIA Jetson Orin Nano}} \\
        \cmidrule(lr){4-6}\cmidrule(lr){7-9}\cmidrule(lr){10-12}
        \textbf{Model} & \textbf{Params} & \textbf{Compr.}
        & \makecell{Lat.\\(ms)} & \makecell{Dyn.\\Pwr (W)} & \makecell{Dyn.\\Energy\\(mJ)}
        & \makecell{Lat.\\(ms)} & \makecell{Dyn.\\Pwr (W)$^\dagger$} & \makecell{Dyn.\\Energy\\(mJ)$^\dagger$}
        & \makecell{Lat.\\(ms)} & \makecell{Dyn.\\Pwr (W)} & \makecell{Dyn.\\Energy\\(mJ)} \\
        \midrule
        SE-ResNet18 & 11.84 M & 1$\times$   & 4.61 & 5.30 & 24.43 & 19.19 & 17.56 & 337.03 & 12.93 & 3.61 & 46.67 \\
        ResNet18   & 11.75 M & --          & 2.82 & 4.79 & 13.53 & 17.64 & 18.91 & 333.48 & 9.5 & 4.92 & 46.74 \\
        SE-ResNet8  & 1.56 M  & 7.6$\times$ & 2.19 & 1.42 & 3.11  & 12.22 & 18.86 & 230.45 & 5.73 & 2.73 & 15.64 \\
        ResNet8    & 1.51 M  & 7.8$\times$ & 1.47 & 1.32 & 1.94  & 26.35 & 20.54 & 541.25 & 4.13 & 3.45 & 14.25 \\
        \bottomrule
    \end{tabular}%
    }
    \vspace{-1mm}
    \raggedright
    \footnotesize{$^\dagger$CPU power and energy are proxy estimates derived from a utilization-scaled TDP model rather than direct device-level telemetry.}
    \vspace{-1mm}
\end{table*}

\begin{figure*}[t]
    \centering
    \includegraphics[width=0.85\linewidth]{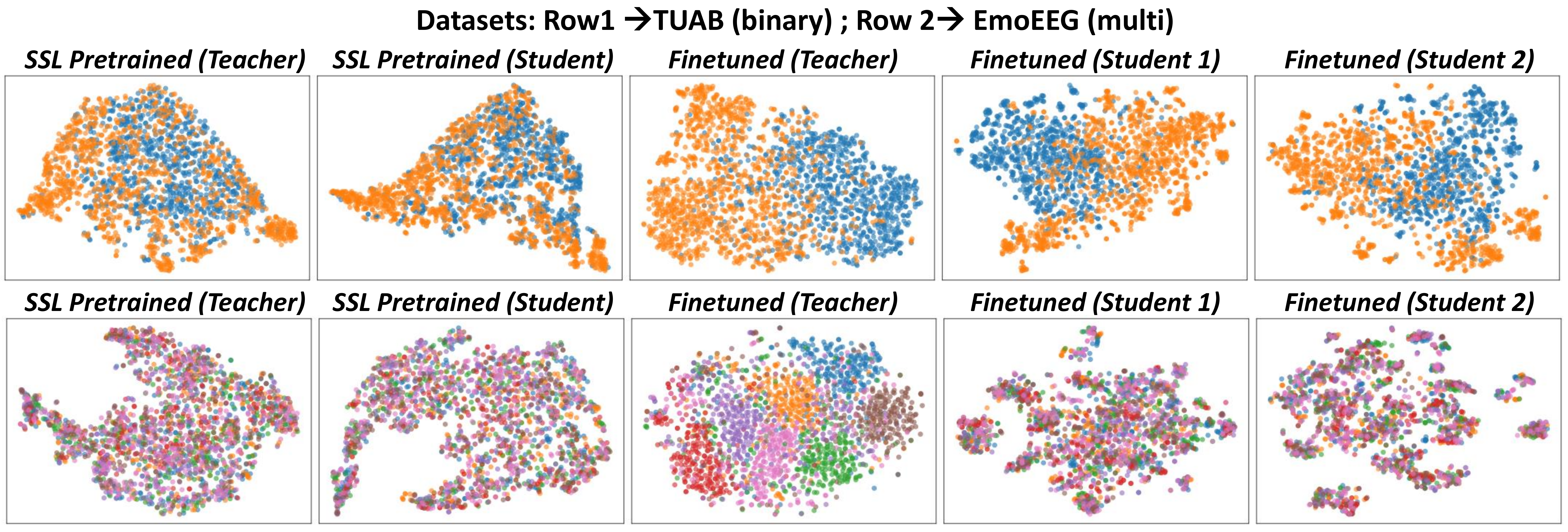}
    ~\vspace{-2mm}
    \caption{t-SNE visualization of BRIDGE-EEG representations on  TUAB (Row 1) and EmoEEG (Row 2), comparing SSL-pretrained teacher and KD student models.}
    \label{fig:tsne}
    ~\vspace{-8mm}
\end{figure*}

\vspace{-2mm}
\subsection{Inference Efficiency Across Deployment Hardware}
\label{subsec:deployment}

% To evaluate deployment practicality, we benchmark energy efficiency across three hardware tiers, namely, a server-class GPU, a desktop CPU, and an edge device. This setup lets us examine how model compression affects latency, power, and energy across increasingly resource-constrained deployment scenarios. Each benchmark is performed at batch size 1 to reflect single-sample, real-time inference. We use 50 warmup inferences before timing and power measurements, followed by 200 runs for latency estimation and 500 runs for power estimation. For the GPU measurements, we first measure the idle power \(P_{\mathrm{idle}}\) which denotes the mean power draw measured after warm-up and before the benchmark loop, average power \(P_{\mathrm{avg}}\) which denotes the mean power draw during inference, and then calculate dynamic power (Eq.~\ref{eq:pdyn}). Under batch size 1, we compute dynamic energy per inference \(E_{\mathrm{dyn}}\) using Eq.~\ref{eq:edyn_latency}, where \(T\) denotes the latency.

To evaluate deployment practicality, we benchmark inference energy across three hardware tiers: a server-class GPU, a desktop CPU, and an edge device, all at batch size 1 to reflect single-sample, real-time inference. We use 50 warmup inferences followed by 200 runs for latency and 500 runs for power estimation. For each platform, we measure idle power $P_{\mathrm{idle}}$ (mean draw before benchmarking) and average inference power $P_{\mathrm{avg}}$ (mean draw during benchmarking), then compute dynamic power as $P_{\mathrm{dyn}} = P_{\mathrm{avg}} - P_{\mathrm{idle}}$ and energy per inference as $E_{\mathrm{dyn}} = P_{\mathrm{dyn}} \times T$, where $T$ is the latency.

\textbf{GPU Tier:}
For the GPU Tier, we perform experiments on the NVIDIA L4. As shown in Table~\ref{tab:deployment}, the SE-ResNet8 student reduces latency from 4.61\,ms to 2.19\,ms and dynamic energy from 24.43\,mJ to 3.11\,mJ relative to the SE-ResNet18 teacher, corresponding to a $7.9\times$ reduction in energy per inference. This shows that the student is not only smaller in parameter count, but also much cheaper to run, which is critical for real-time EEG inference. Comparing SE and non-SE variants, the power gap is negligible but the latency gap is larger, similar to our previous study~\cite{chowdhury2025sslse}. ResNet8 and SE-ResNet8 have similar dynamic power, 1.32\,W versus 1.42\,W, but their latencies differ more clearly, 1.47\,ms versus 2.19\,ms, increasing dynamic energy per inference from 1.94\,mJ to 3.11\,mJ. The same pattern holds at the larger-model scale, where ResNet18 and SE-ResNet18 differ modestly in dynamic power (4.79\,W vs.\ 5.30\,W) but latency increases from 2.82\,ms to 4.61\,ms, raising energy from 13.53\,mJ to 24.43\,mJ. This means the hardware cost of SE is seen more clearly in energy per sample than in power alone. The accuracy benefit of SE blocks over their non-SE counterparts was established in our prior study~\cite{chowdhury2025sslse}, where SE variants consistently improved classification accuracy with negligible power overhead; the present profiling extends that analysis to measured latency and energy across deployment tiers.

\textbf{CPU Tier:}
On the desktop CPU tier (i7-8700), we estimate CPU power using the rated TDP of 65\,W as a baseline. As shown in Table~\ref{tab:deployment}, the SE-ResNet8 student reduces latency from 19.19\,ms for the SE-ResNet18 teacher to 12.22\,ms, indicating that compression improves inference practicality in a desktop setting. Interestingly, the non-SE ResNet8 reaches 26.35\,ms, which is slower than both SE-ResNet8 and the larger ResNet18 at 17.64\,ms. This suggests that realized latency cannot be inferred from parameter count alone, and instead depends on how each architecture maps to the target hardware. As noted in Table~\ref{tab:deployment}, CPU power figures are proxy estimates and may overestimate dynamic power.

\textbf{Edge Tier:} For the edge device tier, we profile the NVIDIA Jetson Orin Nano, which operates under tighter power and memory constraints compared to server and desktop platforms. As shown in Table~\ref{tab:deployment}, the SE-ResNet8 student achieves a latency of 5.73\,ms, dynamic power of 2.73\,W, and dynamic energy of 15.64\,mJ per inference. In comparison, the non-SE ResNet8 achieves a lower latency of 4.13\,ms but higher dynamic power of 3.45\,W, resulting in a comparable dynamic energy of 14.25\,mJ. The SE-ResNet18 teacher incurs significantly higher cost, with 12.93\,ms latency, 3.61\,W dynamic power, and 46.67\,mJ dynamic energy per inference. These results show that model compression leads to substantial improvements in both latency and energy efficiency on edge hardware. While SE blocks introduce a modest latency overhead relative to non-SE variants, they reduce dynamic power and maintain competitive energy consumption. Overall, among the models profiled, SE-ResNet8 offers a favorable accuracy-to-energy tradeoff for edge deployment.

In summary, these results support a deployment pathway in which the teacher is best suited to server-class inference, while SE-ResNet8 provides a stronger accuracy--efficiency tradeoff for desktop and edge deployment.

\vspace{-2mm}
\section{Discussion}
\label{sec:discussion}

% The results show that the proposed framework can produce compact, efficient EEG models across multiple downstream settings. In this section, we discuss how these properties manifest in the learned representations and how they translate into a practical pathway toward wearable deployment.
Here, we examine how the learned representations transfer from teacher to student, and how the resulting compact models relate to a practical pathway toward wearable deployment.
\vspace{-2mm}
\subsection{Knowledge Transfer from Teacher to Student}
To characterize the transfer of learning from pretraining to compact students, we compare the representation spaces of the SSL teacher, KD‑pretrained students, and fine‑tuned models using Centered Kernel Alignment (CKA) (Table~\ref{tab:cka_main}) and t‑SNE visualizations (Fig.~\ref{fig:tsne}) on TUAB and EmoEEG datasets. Across both datasets, the SSL-pretrained student starts extremely close to the SSL-teacher (CKA $~\approx$ 0.85–0.92), indicating that KD Strategy A successfully places the student in almost the same representation regime as the full teacher before any labels are used. Downstream fine‑tuning then reshapes these SSL‑like embeddings in a task‑dependent manner. For the teacher, CKA between SSL and fine‑tuned representations is low ($\approx$ 0.19–0.25), showing that task supervision induces a substantial reorganization of the feature space. Students also move away from their KD-pretrained geometry, but to a lesser extent, and the resulting alignment with the fine‑tuned teacher is moderate on TUAB (CKA $~\approx$ 0.48 for SE‑ResNet8, 0.37 for SE‑ResNet4) and weak on EmoEEG ($\approx$ 0.22–0.24). Together with the t‑SNE plots in Fig.~\ref{fig:tsne}, which show clearer normal/abnormal clusters on TUAB and more entangled emotion classes on EmoEEG, these results suggest a common pattern wherein KD Strategy A first makes the student mimic the SSL teacher’s generic EEG representation, and downstream fine‑tuning with KD Strategy B then pulls both teacher and student toward task‑specific solutions, with small students only partially following the teacher and diverging more on the harder, data‑scarce settings.

\begin{table}[t]
    \centering
    % \footnotesize
    \caption{Linear CKA similarity between model variants}
    \label{tab:cka_main}
    % ~\vspace{-5mm}
    \begin{tabular}{llcc}
        \toprule
        \textbf{Model pair} & \textbf{EmoEEG} & \textbf{TUAB} \\
        \midrule

        SSL SE-ResNet18 vs FT SE-ResNet18
        & 0.1946 & 0.2510 \\
        % \midrule
        
         SSL SE-ResNet18 vs KD-pretrained SE-ResNet8
        & 0.9166 & 0.8467 \\
        % \midrule
        
         FT SE-ResNet18 vs KD SE-ResNet8 FT
        & 0.2410 & 0.4786 \\
        FT SE-ResNet18 vs KD SE-ResNet4 FT
        & 0.2215 & 0.3662 \\
        \bottomrule

    \end{tabular}
\end{table}
~\vspace{-3mm}

\vspace{-3mm}
\subsection{Towards Wearable Deployment}
Building on this representational view, we next discuss how these compact students can plausibly be deployed on actual wearable-class hardware. To connect BRIDGE-EEG to various hardware targets, we relate their parameter footprints to the deployment tiers summarized in Table~\ref{tab:tiers}. At the server and desktop tiers, memory is abundant (16+ GB), so the 11.84\,M-parameter SE-ResNet18 teacher and 1.56\,M-parameter SE-ResNet8 student can be stored and executed without constraint. In contrast, the MCU tier is limited to roughly 256\,kB--1\,MB of on-chip memory, which motivates the 0.48\,M-parameter SE-ResNet4 student as our candidate for the wearable. For context, the most compact variant of FEMBA, a recent Mamba-based EEG FM that explicitly targets edge deployment, has its Tiny model at 7.8\,M parameters ($\approx$30\,MB at fp32), an order of magnitude larger than SE-ResNet4 and well outside the MCU memory envelope~\cite{tegon2025femba}, though the two architectures differ in input representation and are not directly comparable in accuracy. Furthermore, FEMBA reports efficiency in FLOPs rather than measured on-device latency or energy, whereas our Jetson Orin Nano profiling (5.73\,ms, 15.64\,mJ per inference for SE-ResNet8) provides the kind of concrete hardware benchmark that theoretical FLOPs counts cannot substitute. 
With 32-bit weights, SE-ResNet4 requires about $0.48 \times 10^{6} \times 4 \approx 1.9~\text{MB}$ of storage, whereas SE-ResNet8 and SE-ResNet18 require approximately 6.2\,MB and 47\,MB, respectively. Cortex-M-class ultra-low-power microcontrollers (e.g., STM32U5~\cite{STM32U5s91:online}) offer up to 2--4\,MB of on-chip flash and hundreds of kilobytes of SRAM and optional external Quad Serial Peripheral Interface (QSPI) flash, placing SE-ResNet4 within reach for a single-chip or chip-plus-QSPI design, while the larger models would require more aggressive compression. This estimate covers weight storage only. Fitting runtime activations and the input tensor within MCU SRAM would require additional strategies such as layer-wise execution or reduced-precision tiling. With standard 16-bit or 8-bit quantization, the SE-ResNet4 weights would shrink to roughly 0.96\,MB or 0.48\,MB, respectively, leaving headroom for signal-processing code and buffers on a Cortex-M-class wearable node. Furthermore, unlike hybrid convolutional–attention architectures such as EEG Conformer~\cite{song2023conformer}, the purely convolutional design of SE-ResNet exhibits regular memory access patterns that are better suited to standard MCU quantization toolchains and fixed-point execution. Future work will focus on hardware-aware quantization and MCU-class deployment, channel-count reduction for realistic wearable montages, and subject-adaptive training strategies that support practical on-device personalization.

% While EEGConformer achieves good accuracy with fewer parameters, its conv-attention architecture introduces irregular memory access patterns that are difficult to deploy on edge hardware. SE-ResNet8 being pure CNN is quantization-friendly with regular memory access, and serves as a single shared backbone across tasks, making it more practical for wearable deployment. Additionally, EEGConformer is trained from scratch per subject and per dataset, requiring a separate model for each task, and does not generalize across datasets or recording setups. In contrast, our SE-ResNet8 is distilled from a single pretrained backbone and generalizes across six benchmarks spanning three task categories, with only a lightweight classification head swapped per task.

\begin{table}[t]
    \centering
    \footnotesize
    \setlength{\tabcolsep}{3pt}
    \caption{Model family and target deployment tiers.}
    \label{tab:tiers}
    \begin{tabular}{p{1cm}p{2cm}p{1.5cm}p{1.15cm}p{2.2cm}}
    % \begin{tabular}{llcccc}
        \toprule
        \textbf{Tier} & \textbf{Platform} & \textbf{Power} & \textbf{Memory} & \textbf{Model}  \\
        \midrule
        Server  & NVIDIA L4   & 50--300\,W   & 16+\,GB       & Teacher (11.84 M)   \\
        Desktop & i7-8700 CPU & 65\,W TDP    & 16+\,GB       & SE-ResNet8 (1.56 M)   \\
        Edge    & NVIDIA Jetson Orin Nano & 7--15\,W     & 2--8\,GB      & SE-ResNet8 (1.56 M)   \\
        MCU     & Cortex‑M‑class MCU   & 10--100\,mW  & 256\,KB--1\,MB & SE-ResNet4 (0.48 M) \\
        \bottomrule
    \end{tabular}
    ~\vspace{-4mm}
\end{table}

\vspace{-2.2mm}
\section{Conclusion}
\label{sec:conclusion}
This work presents an SSL-to-KD pipeline for compressing a pretrained SE-ResNet18 teacher into compact multi-task EEG models that better balance accuracy and efficiency. Across six downstream benchmarks, the student models retained useful performance after substantial compression on tasks where the teacher's pretrained representation was strong, while on motor imagery the bottleneck was traced to pretraining diversity rather than compression itself. The two distillation strategies served different purposes. Strategy~A produced a shared initialization across tasks, while Strategy~B gave stronger task-specific accuracy, especially for abnormality detection. F1 analysis further revealed that accuracy alone can overstate compression success on class-imbalanced tasks, and that Strategy~A tends to produce more balanced per-class predictions. Inference profiling on server GPU, desktop CPU, and an NVIDIA Jetson Orin Nano edge device showed that compression also reduced latency and energy per inference in practice. Overall, this work provides measured efficiency benchmarks that motivate further compression and MCU-class deployment for wearables.

% supports a practical path from large pretrained EEG models toward compact systems for future low-power and wearable EEG applications.

%% ====================================================================
%%  REFERENCES
%% ====================================================================

\vspace{-3mm}
\section*{References}
~\vspace{-1.2cm}
\footnotesize
% Generated by IEEEtran.bst, version: 1.14 (2015/08/26)


\begin{thebibliography}{10}
\providecommand{\url}[1]{#1}
\csname url@samestyle\endcsname
\providecommand{\newblock}{\relax}
\providecommand{\bibinfo}[2]{#2}
\providecommand{\BIBentrySTDinterwordspacing}{\spaceskip=0pt\relax}
\providecommand{\BIBentryALTinterwordstretchfactor}{4}
\providecommand{\BIBentryALTinterwordspacing}{\spaceskip=\fontdimen2\font plus
\BIBentryALTinterwordstretchfactor\fontdimen3\font minus
  \fontdimen4\font\relax}
\providecommand{\BIBforeignlanguage}[2]{{%
\expandafter\ifx\csname l@#1\endcsname\relax
\typeout{** WARNING: IEEEtran.bst: No hyphenation pattern has been}%
\typeout{** loaded for the language `#1'. Using the pattern for}%
\typeout{** the default language instead.}%
\else
\language=\csname l@#1\endcsname
\fi
#2}}
\providecommand{\BIBdecl}{\relax}
\BIBdecl

\bibitem{he2023diversity}
C.~He \emph{et~al.}, ``Diversity and suitability of the state-of-the-art
  wearable and wireless {EEG} systems review,'' \emph{IEEE J. Biomed. Health
  Inform.}, vol.~27, no.~8, pp. 3830--3843, 2023.

\bibitem{ouahidi2025reve}
Y.~E. Ouahidi \emph{et~al.}, ``{REVE}: A foundation model for {EEG}--adapting
  to any setup with large-scale pretraining on 25,000 subjects,'' \emph{arXiv
  preprint arXiv:2510.21585}, 2025.

\bibitem{wei2025neurolm}
W.~Wei \emph{et~al.}, ``{NeuroLM}: A universal multi-task foundation model for
  bridging the gap between language and {EEG} signals,'' in \emph{Proc. ICLR},
  2025.

\bibitem{kostas2021bendr}
D.~Kostas \emph{et~al.}, ``{BENDR}: Using transformers and a contrastive
  self-supervised learning task to learn from massive amounts of {EEG} data,''
  \emph{Front. Hum. Neurosci.}, vol.~15, p. 653659, 2021.

\bibitem{zhou2025csbrain}
Y.~Zhou \emph{et~al.}, ``{CSBrain}: A cross-scale spatiotemporal brain
  foundation model for {EEG} decoding,'' \emph{arXiv preprint
  arXiv:2506.23075}, 2025.

\bibitem{yang2023biot}
C.~Yang \emph{et~al.}, ``{BIOT}: Cross-data biosignal learning in the wild,''
  in \emph{Proc. NeurIPS}, 2023.

\bibitem{wang2025cbramod}
J.~Wang \emph{et~al.}, ``{CBraMod}: A criss-cross brain foundation model for
  {EEG} decoding,'' in \emph{Proc. ICLR}, 2025.

\bibitem{jiang2024labram}
W.~Jiang \emph{et~al.}, ``Large brain model for learning generic
  representations with tremendous {EEG} data in {BCI},'' in \emph{Proc. ICLR},
  2024, spotlight.

\bibitem{panchavati2025mentality}
S.~Panchavati \emph{et~al.}, ``Mentality: A {Mamba}-based approach towards
  foundation models for {EEG},'' \emph{arXiv preprint arXiv:2509.02746}, 2025.

\bibitem{wang2025eegmamba}
J.~Wang \emph{et~al.}, ``{EEGMamba}: An {EEG} foundation model with {Mamba},''
  \emph{Neural Networks}, p. 107816, 2025.

\bibitem{tegon2025femba}
A.~Tegon \emph{et~al.}, ``{FEMBA}: Efficient and scalable {EEG} analysis with a
  bidirectional {Mamba} foundation model,'' in \emph{2025 47th Annual
  International Conference of the IEEE Engineering in Medicine and Biology
  Society (EMBC)}.\hskip 1em plus 0.5em minus 0.4em\relax IEEE, 2025, pp. 1--7.

\bibitem{chen2020simclr}
T.~Chen \emph{et~al.}, ``A simple framework for contrastive learning of visual
  representations,'' in \emph{Proc. ICML}, 2020.

\bibitem{chowdhury2025sslse}
M.~R. Chowdhury \emph{et~al.}, ``{SSL-SE-EEG}: A framework for robust learning
  from unlabeled {EEG} data with self-supervised learning and
  squeeze-excitation networks,'' in \emph{Proc. IEEE EMBC}, 2025,
  arXiv:2510.19829.

\bibitem{wang2023brainbert}
C.~Wang \emph{et~al.}, ``{BrainBERT}: Self-supervised representation learning
  for intracranial recordings,'' in \emph{Proc. ICLR}, 2023.

\bibitem{zhang2023brant}
Z.~Zhang \emph{et~al.}, ``Brant: Foundation model for intracranial neural
  signal,'' in \emph{Proc. NeurIPS}, 2023.

\bibitem{cui2024neurogpt}
W.~Cui \emph{et~al.}, ``{NeuroGPT}: Towards a foundation model for {EEG},'' in
  \emph{Proc. IEEE ISBI}, 2024.

\bibitem{wang2024eegpt}
G.~Wang \emph{et~al.}, ``{EEGPT}: Pretrained transformer for universal and
  reliable representation of {EEG} signals,'' \emph{Advances in Neural
  Information Processing Systems}, vol.~37, pp. 39\,249--39\,280, 2024.

\bibitem{lawhern2018eegnet}
V.~J. Lawhern \emph{et~al.}, ``{EEGNet}: A compact convolutional neural network
  for {EEG}-based brain-computer interfaces,'' \emph{J. Neural Eng.}, vol.~15,
  no.~5, p. 056013, 2018.

\bibitem{schirrmeister2017deep}
R.~T. Schirrmeister \emph{et~al.}, ``Deep learning with convolutional neural
  networks for {EEG} decoding and visualization,'' \emph{Hum. Brain Mapp.},
  vol.~38, no.~11, pp. 5391--5420, 2017.

\bibitem{hu2018senet}
J.~Hu \emph{et~al.}, ``Squeeze-and-excitation networks,'' in \emph{Proc. IEEE
  CVPR}, 2018, pp. 7132--7141.

\bibitem{li2020tsseseizure}
Y.~Li \emph{et~al.}, ``Epileptic seizure detection in {EEG} signals using a
  unified temporal-spectral squeeze-and-excitation network,'' \emph{IEEE Trans.
  Neural Syst. Rehabil. Eng.}, vol.~28, no.~4, pp. 782--794, 2020.

\bibitem{altuwaijri2022mbeegse}
G.~A. Altuwaijri \emph{et~al.}, ``A multi-branch convolutional neural network
  with squeeze-and-excitation attention blocks for {EEG}-based motor imagery
  signals classification,'' \emph{Diagnostics}, vol.~12, no.~4, p. 995, 2022.

\bibitem{ingolfsson2020eegtcnet}
T.~M. Ingolfsson \emph{et~al.}, ``{EEG-TCNet}: An accurate temporal
  convolutional network for embedded motor-imagery brain-machine interfaces,''
  in \emph{Proc. IEEE SMC}, 2020, pp. 2958--2965.

\bibitem{song2023conformer}
Y.~Song \emph{et~al.}, ``{EEG Conformer}: Convolutional transformer for {EEG}
  decoding and visualization,'' \emph{IEEE Trans. Neural Syst. Rehabil. Eng.},
  vol.~31, pp. 710--719, 2023.

\bibitem{altaheri2023atcnet}
H.~Altaheri \emph{et~al.}, ``Physics-informed attention temporal convolutional
  network for {EEG}-based motor imagery classification,'' \emph{IEEE Trans.
  Ind. Inform.}, vol.~19, no.~2, pp. 2249--2258, 2023.

\bibitem{banville2021structure}
H.~Banville \emph{et~al.}, ``Uncovering the structure of clinical {EEG} signals
  with self-supervised learning,'' \emph{J. Neural Eng.}, vol.~18, no.~4, p.
  046003, 2021.

\bibitem{mohsenvand2020contrastive}
M.~N. Mohsenvand \emph{et~al.}, ``Contrastive representation learning for
  electroencephalogram classification,'' in \emph{ML4H Workshop, NeurIPS (PMLR
  vol.~136)}, 2020, pp. 238--253.

\bibitem{chien2022maeeg}
H.-Y.~S. Chien \emph{et~al.}, ``{MAEEG}: Masked auto-encoder for {EEG}
  representation learning,'' \emph{arXiv:2211.02625}, 2022.

\bibitem{foumani2024eeg2rep}
N.~M. Foumani \emph{et~al.}, ``{EEG2Rep}: Enhancing self-supervised {EEG}
  representation through informative masked inputs,'' in \emph{Proc. KDD},
  2024, pp. 5544--5555.

\bibitem{xiao2024slamseizure}
T.~Xiao \emph{et~al.}, ``Self-supervised learning with attention mechanism for
  {EEG}-based seizure detection,'' \emph{Biomed. Signal Process. Control},
  2024.

\bibitem{li2024sslmi}
W.~Li \emph{et~al.}, ``Self-supervised contrastive learning for {EEG}-based
  cross-subject motor imagery recognition,'' \emph{J. Neural Eng.}, vol.~21,
  no.~2, 2024.

\bibitem{hinton2015distilling}
G.~Hinton \emph{et~al.}, ``Distilling the knowledge in a neural network,''
  \emph{arXiv:1503.02531}, 2015.

\bibitem{romero2015fitnets}
A.~Romero \emph{et~al.}, ``{FitNets}: Hints for thin deep nets,'' in
  \emph{Proc. ICLR}, 2015.

\bibitem{huang2023skd}
X.-Y. Huang \emph{et~al.}, ``Enhancing low-density {EEG}-based brain-computer
  interfaces with similarity-keeping knowledge distillation,'' \emph{IEEE
  Trans. Emerg. Topics Comput. Intell.}, vol.~8, no.~2, pp. 1156--1166, 2023.

\bibitem{liang2023sleepkd}
H.~Liang \emph{et~al.}, ``Teacher assistant-based knowledge distillation
  extracting multi-level features on single channel sleep {EEG},'' in
  \emph{Proc. IJCAI}, 2023, pp. 3948--3956.

\bibitem{jia2024distillsleep}
Z.~Jia \emph{et~al.}, ``{DistillSleepNet}: Heterogeneous multi-level knowledge
  distillation via teacher assistant for sleep staging,'' \emph{IEEE Trans. Big
  Data}, vol.~11, pp. 1273--1284, 2024.

\bibitem{zhang2022multichannel}
C.~Zhang \emph{et~al.}, ``Multi-channel multi-domain based knowledge
  distillation algorithm for sleep staging with single-channel {EEG},''
  \emph{IEEE TCAS-II}, vol.~69, no.~11, pp. 4608--4612, 2022.

\bibitem{zhang2024llmkd}
Y.~Zhang \emph{et~al.}, ``{LLM}-enhanced multi-teacher knowledge distillation
  for modality-incomplete emotion recognition in daily healthcare,'' \emph{IEEE
  J. Biomed. Health Inform.}, vol.~29, no.~9, pp. 6406--6416, 2024.

\bibitem{zhang2022visualeeg}
S.~Zhang \emph{et~al.}, ``Visual-to-{EEG} cross-modal knowledge distillation
  for continuous emotion recognition,'' \emph{Pattern Recognit.}, vol. 130, p.
  108833, 2022.

\bibitem{baghersalimi2024m2skd}
S.~Baghersalimi \emph{et~al.}, ``{M2SKD}: Multi-to-single knowledge
  distillation of real-time epileptic seizure detection for low-power wearable
  systems,'' \emph{ACM Trans. Intell. Syst. Technol.}, vol.~15, no.~5, pp.
  1--31, 2024.

\bibitem{sepahvand2022ecgkd}
M.~Sepahvand and F.~Abdali-Mohammadi, ``A novel method for reducing arrhythmia
  classification from 12-lead {ECG} signals to single-lead {ECG} with minimal
  loss of accuracy through teacher-student knowledge distillation,'' \emph{Inf.
  Sci.}, vol. 593, pp. 64--77, 2022.

\bibitem{abbaspourazad2024wearablekd}
S.~Abbaspourazad \emph{et~al.}, ``Wearable accelerometer foundation models for
  health via knowledge distillation,'' \emph{arXiv:2412.11276}, 2024.

\bibitem{ingolfsson2024brainfusenet}
T.~M. Ingolfsson \emph{et~al.}, ``{BrainFuseNet}: Enhancing wearable seizure
  detection through {EEG-PPG}-accelerometer sensor fusion and efficient edge
  deployment,'' \emph{IEEE Trans. Biomed. Circuits Syst.}, vol.~18, no.~4, pp.
  720--733, 2024.

\bibitem{ingolfsson2023epidenet}
------, ``{EpiDeNet}: An energy-efficient approach to seizure detection for
  embedded systems,'' in \emph{2023 IEEE Biomedical Circuits and Systems
  Conference (BioCAS)}.\hskip 1em plus 0.5em minus 0.4em\relax IEEE, 2023, pp.
  1--5.

\bibitem{zanetti2020stm32}
R.~Zanetti \emph{et~al.}, ``Robust epileptic seizure detection on wearable
  systems with reduced false-alarm rate,'' in \emph{Proc. IEEE EMBC}, 2020, pp.
  4248--4251.

\bibitem{saric2020fpga}
R.~Sari\'{c} \emph{et~al.}, ``{FPGA}-based real-time epileptic seizure
  classification using artificial neural network,'' \emph{Biomed. Signal
  Process. Control}, vol.~62, p. 102106, 2020.

\bibitem{li2022asic65nm}
H.~Li \emph{et~al.}, ``A {65nm/0.448\,mW} {EEG} processor with parallel
  architecture {SVM} and lifting wavelet transform for high-performance and
  low-power epilepsy detection,'' \emph{Comput. Biol. Med.}, 2022.

\bibitem{jurcak200710}
V.~Jurcak \emph{et~al.}, ``10/20, 10/10, and 10/5 systems revisited: their
  validity as relative head-surface-based positioning systems,''
  \emph{Neuroimage}, vol.~34, no.~4, pp. 1600--1611, 2007.

\bibitem{harris1978windows}
F.~J. Harris, ``On the use of windows for harmonic analysis with the discrete
  {F}ourier transform,'' \emph{Proceedings of the IEEE}, vol.~66, no.~1, pp.
  51--83, 1978.

\bibitem{howard2018ulmfit}
J.~Howard and S.~Ruder, ``Universal language model fine-tuning for text
  classification,'' in \emph{Proceedings of ACL}, 2018, pp. 328--339.

\bibitem{jing2023development}
J.~Jing \emph{et~al.}, ``Development of expert-level classification of seizures
  and rhythmic and periodic patterns during {EEG} interpretation,''
  \emph{Neurology}, vol. 100, no.~17, pp. e1750--e1762, 2023.

\bibitem{yang2023self}
C.~Yang \emph{et~al.}, ``Self-supervised electroencephalogram representation
  learning for automatic sleep staging: model development and evaluation
  study,'' \emph{JMIR AI}, vol.~2, no.~1, p. e46769, 2023.

\bibitem{peh2022transformer}
W.~Y. Peh \emph{et~al.}, ``Transformer convolutional neural networks for
  automated artifact detection in scalp {EEG},'' in \emph{2022 44th Annual
  International Conference of the IEEE Engineering in Medicine \& Biology
  Society (EMBC)}.\hskip 1em plus 0.5em minus 0.4em\relax IEEE, 2022, pp.
  3599--3602.

\bibitem{li2022motor}
H.~Li \emph{et~al.}, ``Motor imagery {EEG} classification algorithm based on
  {CNN-LSTM} feature fusion network,'' \emph{Biomedical signal processing and
  control}, vol.~72, p. 103342, 2022.

\bibitem{jiang2023epileptic}
X.~Jiang \emph{et~al.}, ``Epileptic seizures detection and the analysis of
  optimal seizure prediction horizon based on frequency and phase analysis,''
  \emph{Frontiers in Neuroscience}, vol.~17, p. 1191683, 2023.

\bibitem{wang2023automatic}
Z.~Wang \emph{et~al.}, ``Automatic epileptic seizure detection based on
  persistent homology,'' \emph{Frontiers in physiology}, vol.~14, p. 1227952,
  2023.

\bibitem{yang2023patient}
Y.~Yang \emph{et~al.}, ``Patient-specific approach using data fusion and
  adversarial training for epileptic seizure prediction,'' \emph{Frontiers in
  Computational Neuroscience}, vol.~17, p. 1172987, 2023.

\bibitem{STM32U5s91:online}
\BIBentryALTinterwordspacing
``{STM32U5} series of ultra-low-power {MCU}s enhanced security for {IoT} and
  embedded applications - {STMicroelectronics},'' [Online; accessed
  2026-03-27]. [Online]. Available:
  \url{https://www.st.com/en/microcontrollers-microprocessors/stm32u5-series.html}
\BIBentrySTDinterwordspacing

\end{thebibliography}
\end{document}